\documentclass[a4paper,usenatbib]{mnras}

\usepackage[T1]{fontenc}
\usepackage{ae,aecompl}

\usepackage{graphicx}	
\usepackage{amsmath}	
\usepackage{amssymb}	
\usepackage{textcomp}
\usepackage{upgreek}
\usepackage{hyperref}
\usepackage{breakurl}
\usepackage{soul}

\newcommand{\urlwofont}[1]{\urlstyle{same}\url{#1}}
\newcommand{\kms}{km~s$^{-1}$}

\title[SN~1987A ejecta morphology]{The morphology of the ejecta of SN~1987A at 31 years from 1150 to 10000 \AA}

\author[T. Kangas et al.]{T. Kangas,$^{1}$\thanks{E$-$mail: tuomask@kth.se}
C. Fransson,$^{2}$ 
J. Larsson,$^{1}$
K. France,$^{3,4,5}$
R. A. Chevalier,$^{6}$
R. P. Kirshner,$^{7,8}$ \newauthor
P. Lundqvist,$^{2}$
S. Mattila,$^{9}$
J. Sollerman$^{2}$
and V. P. Utrobin,$^{10,11,12}$
\\
$^{1}$Department of Physics, KTH Royal Institute of Technology, The Oskar Klein Centre, AlbaNova, SE-106 91 Stockholm, Sweden\\
$^{2}$Department of Astronomy, Stockholm University, The Oskar Klein Centre, AlbaNova, SE-106 91 Stockholm, Sweden \\
$^{3}$Laboratory for Atmospheric and Space Physics, University of Colorado, 600 UCB, Boulder, CO 80309 \\
$^{4}$Department of Astrophysical and Planetary Science, University of Colorado, 389 UCB, Boulder, CO 80309, USA \\
$^{5}$Center for Astrophysics and Space Astronomy, University of Colorado, 593 UCB, Boulder, CO 80309 \\
$^{6}$Department of Astronomy, University of Virginia, P.O. Box 400325, Charlottesville, VA 22904-4325, USA \\
$^{7}$Harvard-Smithsonian Center for Astrophysics, 60 Garden Street, Cambridge, MA 02138, USA \\
$^{8}$Gordon and Betty Moore Foundation, 1661 Page Mill Road, Palo Alto, CA 94308 \\
$^{9}$Tuorla observatory, Department of Physics and Astronomy, University of Turku, FI-20014 Turku, Finland \\
$^{10}$NRC ‘Kurchatov Institute’ – Institute for Theoretical and Experimental Physics, B. Cheremushkinskaya St. 25, 117218 Moscow, Russia \\
$^{11}$Institute of Astronomy, Russian Academy of Sciences, Pyatnitskaya St. 48, 119017 Moscow, Russia \\
$^{12}$Max-Planck-Institut f\"{u}r Astrophysik, Karl-Schwarzschild-Str. 1, 85748 Garching, Germany \\
}
\date{Accepted XXX. Received YYY; in original form ZZZ}
\pubyear{2021}

\begin{document}
\label{firstpage}
\pagerange{\pageref{firstpage}$-$\pageref{lastpage}}
\maketitle

\begin{abstract}

We present spectroscopy of the ejecta of SN~1987A in 2017 and 2018 from the \textit{Hubble Space Telescope} and the Very Large Telescope, covering the wavelength range between $1150$ and $10000$~\AA. At 31 years, this is the first epoch with coverage over the ultraviolet-to-near-infrared range since 1995. We create velocity maps of the ejecta in the H$\alpha$, Mg~{\sc ii}~$\lambda\lambda2796,2804$ and [O~{\sc i}]~$\lambda\lambda6302,6366$ (vacuum) emission lines and study their morphology. All three lines have a similar morphology, but Mg~{\sc ii} is blueshifted by $\sim$1000~km~s$^{-1}$ relative to the others and stronger in the northwest. We also study the evolution of the line fluxes, finding a brightening by a factor of $\sim$9 since 1999 in Mg~{\sc ii}, while the other line fluxes are similar in 1999 and 2018. We discuss implications for the power sources of emission lines at late times: thermal excitation due to heating by the X-rays from the ejecta-ring interaction is found to dominate the ultraviolet Mg~{\sc ii} lines, while the infrared Mg~{\sc ii} doublet is powered mainly by Ly$\alpha$ fluorescence. The X-ray deposition is calculated based on merger models of SN 1987A. Far-ultraviolet emission lines of H$_2$ are not detected. Finally, we examine the combined spectrum of recently-discovered hotspots outside the equatorial ring. Their unresolved Balmer emission lines close to zero velocity are consistent with the interaction of fast ejecta and a clumpy, slowly moving outflow. A clump of emission in this spectrum, south of the equatorial ring at $\sim$1500~km~s$^{-1}$, is likely associated with the reverse shock.  

\end{abstract}

\begin{keywords}
supernovae: individual: SN~1987A $-$ ISM: supernova remnants
\end{keywords}


 
\section{Introduction}

The closest directly observed supernova (SN) to us since the 17th century, SN~1987A was discovered on February 24 1987 in the Large Magellanic Cloud (LMC), located at a distance of only $\sim$50 kpc \citep{pietrzynski19}. Its proximity and age have allowed a follow-up of unpredecented detail and timespan, making SN~1987A the best available object for studying various aspects of the deaths of massive stars as core-collapse SNe (CCSNe). These topics include nucleosynthesis, the radioactive decay of $^{56}$Ni, $^{57}$Ni and $^{44}$Ti as the power source, and the evolution of the SN into a supernova remnant (SNR); for insights on the latter, see \citet{milfes17}. In addition to the proximity of the SN itself, the capabilities of the \textit{Hubble Space Telescope} (\textit{HST}) have greatly aided this follow-up, together making it possible to resolve and study emission from SN 1987A in much more detail than in extragalactic SNe. Despite being a rare peculiar Type II (hydrogen-rich) SN with a blue supergiant (BSG) progenitor of $\sim16-22 M_\odot$ \citep[see e.g.][for a discussion and references]{mccray16}, SN~1987A helped confirm theories of CCSNe as transitions to a neutron star through observations of neutrinos from the explosion \citep[e.g.][]{hirata87}, and the follow-up has helped constrain explosion mechanisms and nucleosynthesis through e.g. spectral modeling \citep{jerkstrand11}. The surprisingly large amount of dust created by SN~1987A also challenges current models \citep{matsuura15}. For a review of the insights gained in the decades of observation, see \citet{mccray16}. The neutron star itself has not definitively been found yet due to high extinction by the ejecta \citep{orlando15,alp18}, although some authors \citep{zanardo14,cigan19,greco21} have reported signs of a possible neutron star remnant \citep[but see also][]{alp21}.

\begin{figure}
\centering
\includegraphics[width=\columnwidth]{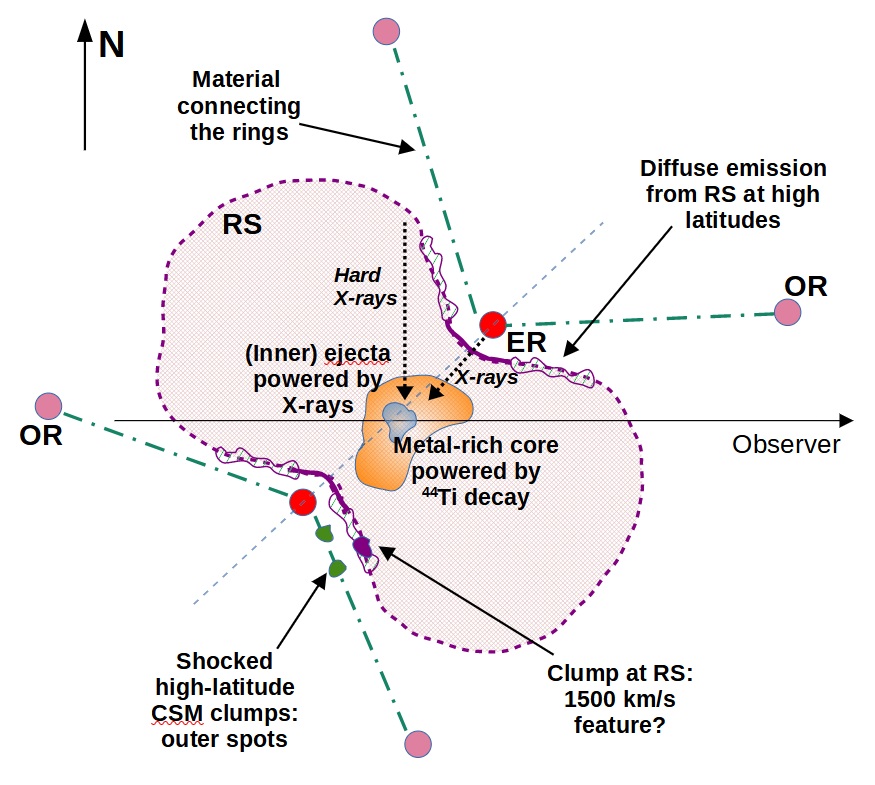}
\caption{Side-view sketch of SN 1987A \citep[based on our results below and ][]{chevalier95,france15,larsson19}. The main optically emitting regions are labeled. The clumps south of the ER correspond to the suggested origin of the hotspots we observe; see Sect. \ref{sec:spotveloc}. The inner ejecta (orange) is where edge-brightened distributions of emission originate, while centrally peaked distributions correspond to the core.}
\label{fig:sketch}
\end{figure}

The evolving remnant of SN~1987A now consists of the ejecta interacting with the circumstellar medium (CSM). The CSM includes diffuse gas, the equatorial ring (ER) $\sim0.6$ ly in radius, and the outer rings (ORs), each $\sim1.3$ ly in radius; the nearly circular rings all have inclinations around $40^\circ$ \citep{tziamtzis11}. Based on the velocities of matter in the rings, they were ejected around 20000 years ago \citep{crotts00}, probably after the merger of a binary system with masses on the order of $15~M_\odot$ and $5~M_\odot$, before which the primary was a red supergiant (RSG) star \citep{mp07,mp09}. \citet{utrobin21} found a numerical SN model of such a merger remnant progenitor \citep[with initial masses of $15~M_\odot$ and $7~M_\odot$, based on][]{menonheger17} a better match to the observables of SN~1987A than single-star models. \citet{orlando20} and \citet{ono20} performed extensive magnetohydrodynamic modelling and found a similar merger scenario \citep[$14~M_\odot$ + $9~M_\odot$;][]{urushibata18} to match the observed properties of the BSG progenitor, the SN itself and the evolution of its remnant. 

The interaction between the ejecta and the ER -- with estimated masses of $17 M_\odot$ and $0.06 M_\odot$, respectively \citep{lf96,mattila10,orlando15} -- was responsible for a brightening in X-rays and the appearance of a few dozen optical hotspots in the ER from the late 1990s onward \citep{borkowski97,sonneborn98, lawrence00} as the forward shock swept into relatively dense clumps of matter, and their subsequent fading since 2009 is a sign that the shocks have swept past the ER and the interaction is slowly coming to an end \citep{fransson15,arendt16}. The ER may now be in the process of being destroyed by the interaction, but \citet{orlando19} have predicted it to survive the interaction, at least until 40 years post-explosion where their simulation ends. The outer ejecta, advancing beyond the ER, are now interacting with the diffuse gas between the ER and the ORs \citep{larsson19}, creating new, fainter hotspots there. A reverse shock (RS) can also be observed, propagating inward in the frame of the ejecta, with a shape suggested to be bipolar with a narrower 'waist' inside the ER \citep{chevalier95,france15}. We summarize the structure of the remnant in Fig. \ref{fig:sketch}.

The three-dimensional structure of the ejecta, an important clue as to the geometry of the explosion itself \citep[e.g.][]{orlando15,wongw15}, has been studied previously by e.g. \citet{kjaer10} and \citet{larsson13,larsson16,larsson19b}. The structure was shown to be asymmetric, its emission predominantly blueshifted in the northern part and redshifted in the south \citep{kjaer10} in a `broken dipole' \citep{larsson16}. The evolving, edge-brightened morphology of the H$\alpha$ line from the inner (unshocked) ejecta was shown to indicate that its power source has changed from $^{44}$Ti decay to X-ray heating from the ongoing interaction between the ER and the ejecta \citep{fransson13,larsson16}. Although the soft X-ray flux peaked in 2014, the interaction site has continued to brighten in hard X-rays until at least 2020 \citep[][]{frank16,alp21,sun21}. This power source is consistent with the increase in optical emission from the ejecta, previously powered by the decay of $^{56}$Ni, $^{57}$Ni and $^{44}$Ti, simultaneously with the emergence of the hotspots \citep{larsson11}. Meanwhile, the power source of the [Si~{\sc i}] and [Fe~{\sc ii}] infrared lines was determined to still be radioactive heating due to their unchanging, centrally peaked emission morphology. Molecular hydrogen (H$_2$) emission was also found to be concentrated more centrally than H$\alpha$, indicating $^{44}$Ti as its power source \citep{fransson16}. It is, however, located in regions of weak emission from dust and from CO and SiO molecules \citep{larsson19b} -- see also \citet[][]{indebetouw14} and \citet{abellan17} -- which suggests the H$_2$ molecules are formed in gas, not on dust grains. 

In this paper, we present new spectroscopic data of the ejecta and ER of SN~1987A from the \textit{HST}, obtaining the first simultaneous UV-to-optical spectrum of the ejecta since 1995 \citep{chugai97} and the first such spectrum of the ER since 1999 \citep{pun02}, as well as the first spectrum of the outer hotspots. We use the spectrum to study the evolution of the fluxes of emission lines. As the \textit{HST} provides enough angular resolution and sensitivity to study different parts of the ejecta separately, we follow the example of \citet{larsson16}, attempting to study the three-dimensional morphology of the ejecta using the H$\alpha$, Mg~{\sc ii}~$\lambda\lambda2796,2804$ and [O~{\sc i}]~$\lambda\lambda6302,6366$ lines. The differences in structure between these lines can provide clues as to the nature of the mechanisms powering the late-time emission. We describe our observations and data reduction in Sect.~\ref{sec:data} and our analysis and results in Sect.~\ref{sec:res}. We discuss the implications of our results in Sect.~\ref{sec:disco} and finally present our conclusions in Sect.~\ref{sec:concl}. More analysis of our data will be presented in another paper, concentrating on the evolution of the RS (Fransson et al., in preparation). 

\begin{table*}
\begin{minipage}{0.67\linewidth}
\centering
\begin{small}
\caption{Log of our \textit{HST} observations.}
\begin{tabular}{lcccccc}
\hline
UT date & Epoch & Instrument & Grating & Aperture & Position\footnote{As depicted in Fig.~\ref{fig:ha_img}.} & Exp. time\\
(YYYY-MM-DD) & (d) & & & & & (s) \\
 \hline
 2017-02-26.7 & 10961 & COS & G130M & PSA & - & 8180 \\
 2017-02-26.9 & 10961 & COS & G160M & PSA & - & 11209 \\
 2018-01-25.9 & 11294 & STIS & G140L & 52X0.2 & 2 & 8856 \\
 2018-01-26.9 & 11295 & STIS & G230L & 52X0.2 & 1 & 2396 \\
 2018-01-26.9 & 11295 & STIS & G230L & 52X0.2 & 2 & 3230 \\
 2018-01-27.0 & 11295 & STIS & G230L & 52X0.2 & 3 & 3230 \\
 2018-01-27.1 & 11295 & STIS & G430L & 52X0.2 & 1 & 2400 \\
 2018-01-27.2 & 11295 & STIS & G430L & 52X0.2 & 2 & 1792 \\
 2018-01-27.2 & 11295 & STIS & G430L & 52X0.2 & 3 & 3384 \\
 2018-01-28.7 & 11297 & STIS & G750L & 52X0.2 & 1 & 2388 \\
 2018-01-28.9 & 11297 & STIS & G750L & 52X0.2 & 2 & 1780 \\
 2018-01-28.8 & 11297 & STIS & G750L & 52X0.2 & 3 & 3360 \\
 \hline
\end{tabular}
\label{tab:obslog_hst}
\end{small}
\end{minipage}
\end{table*}

\begin{table*}
\begin{minipage}{0.99\linewidth}
\centering
\begin{small}
\caption{Description of previously observed spectra used in our analysis.
}
\begin{tabular}{lccccccc}
\hline
UT date & Mean epoch & Instrument & Wavelength range & Spatial extent & PA & Reference \\
(YYYY-MM-DD) & (d) & & (\AA) & & ($\deg$) & \\
 \hline
 2017-10-19 --- 11-22 & 11236 & VLT/UVES & 3200--10000 & Full ejecta + ER & 30 & \citet{larsson19} \\
 2014-08-16 --- 21 & 10037 & \textit{HST}/STIS & 5400--10000 & Ejecta ($0\farcs5$) & 0 & \citet{larsson16} \\
 2011-02-11 --- 03-14 & 8769 & \textit{HST}/COS & 1140--1800 & Full ejecta + ER & - & \citet{france11} \\
 2004-07-18 --- 23 & 6360 & \textit{HST}/STIS & 5400--10000 & Ejecta ($0\farcs6$) & 180 & \citet{heng06} \\
 1999-09-17 & 4588 & \textit{HST}/STIS & 1600--3100 & Ejecta ($0\farcs5$) & 211 & -- \\
 1999-02-21 --- 27 & 4381 & \textit{HST}/STIS & 5300--10200 & Ejecta ($0\farcs5$) & 26 & \citet{larsson13} \\
 1997-10-08 & 3880 & \textit{HST}/STIS & 1600--3100 & Ejecta ($0\farcs5$) & 223 & \citet{1997IAUC.6710....2G} \\
 1997-01-05 & 3604 & \textit{HST}/FOS & 2000--6800 & Full ejecta & - & - \\
 1995-01-07 & 2875 & \textit{HST}/FOS & 1800--8500 & Full ejecta & - & \citet{chugai97}\\
 \hline
\end{tabular}
\label{tab:obs_others}
\end{small}
\end{minipage}
\end{table*}

\section{Observations and data reduction}
\label{sec:data}

\begin{figure}
\centering
\includegraphics[width=\columnwidth]{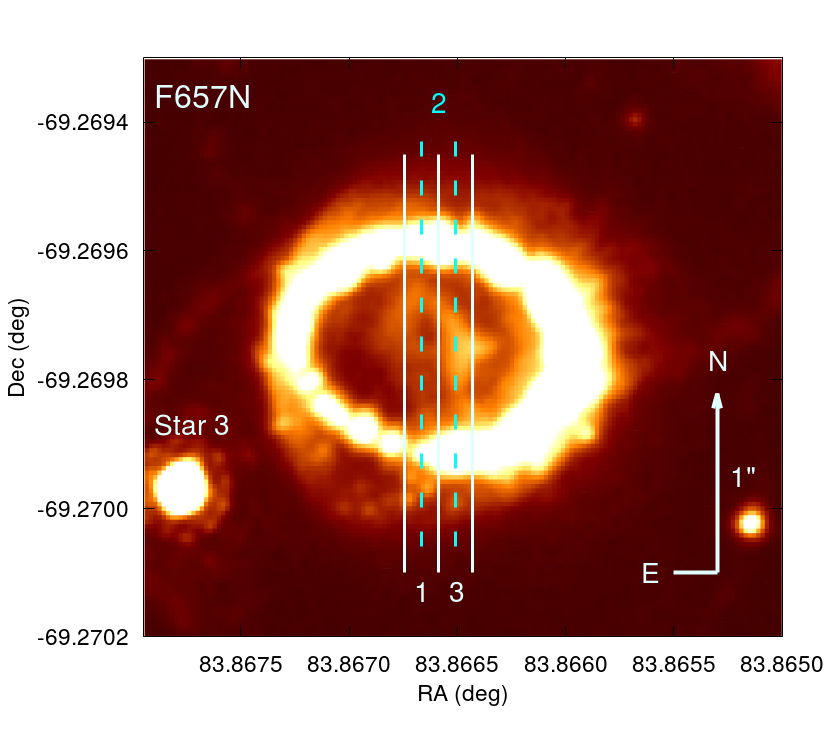}
\caption{Slit positions, referred to as Positions 1, 2 and 3, used in this study. Position 2 (cyan) overlaps with the western half of Position 1 and the eastern half of Position 3. Each slit is 0.2" wide. The western and eastern edges of the ejecta are not covered by any slit. The image was taken using \textit{HST}/WFC3 and the F657N filter on 2018-07-10 \citep{larsson19}.}
\label{fig:ha_img}
\end{figure}

Spectroscopic observations of the ejecta and the ER were performed using the Cosmic Origins Spectrograph (COS) on the \textit{HST} on 2017-02-26 and using the Space Telescope Imaging Spectrograph (STIS), also on the \textit{HST}, between 2018-01-25 and 2018-01-28. On COS, we used the gratings G130M and G160M, covering the wavelength range between 1138 and 1795~\AA; on STIS, we used G140L, G230L, G430L and G750L for a wavelength coverage between 1150 and 10000~\AA. A 0\farcs2 slit, oriented north to south, was used for the STIS observations, while the COS observations were done using the circular 2\farcs5-diameter Primary Science Aperture (PSA). The observations are summarized in Table~\ref{tab:obslog_hst}. These observations were executed as part of \textit{HST} program GO 14753 (PI Fransson). Other data we use for comparison with ours throughout this paper are summarized in Table~\ref{tab:obs_others}. The VLT/UVES spectra from 2017-10-19 are described in \citet{larsson19}.

The angular resolution achievable with STIS ($\sim$0\farcs1) allows us to probe different regions of the ejecta. In the north-south direction, along the slit, it is straightforward to divide the ejecta into multiple extraction regions; in the east-west direction we obtained the spectrum at three positions as shown in Fig.~\ref{fig:ha_img}, in all gratings but G140L, where only the central position (labelled 2) was used. Due to a mistake in the preparation of the observations, the slits were placed 0\farcs1 apart rather than 0\farcs2. As a result, the central slit overlaps with the other two, and the eastern and western edges of the ejecta (as seen in H$\alpha$) are not covered by our observations. Furthermore, instead of the intended three exposures per position, the eastern slit (Position 3) received four exposures while the central slit (Position 2) only received two in our G430L and G750L observations. Meanwhile, COS provides a much better spectral resolution, but its \textit{angular} resolution, with its 2\farcs5 diameter aperture, is not good enough to distinguish between different parts of the ejecta, and the entire ER-ejecta system is included.

\begin{figure*}
\begin{minipage}{\linewidth}
\centering
\includegraphics[width=0.95\columnwidth]{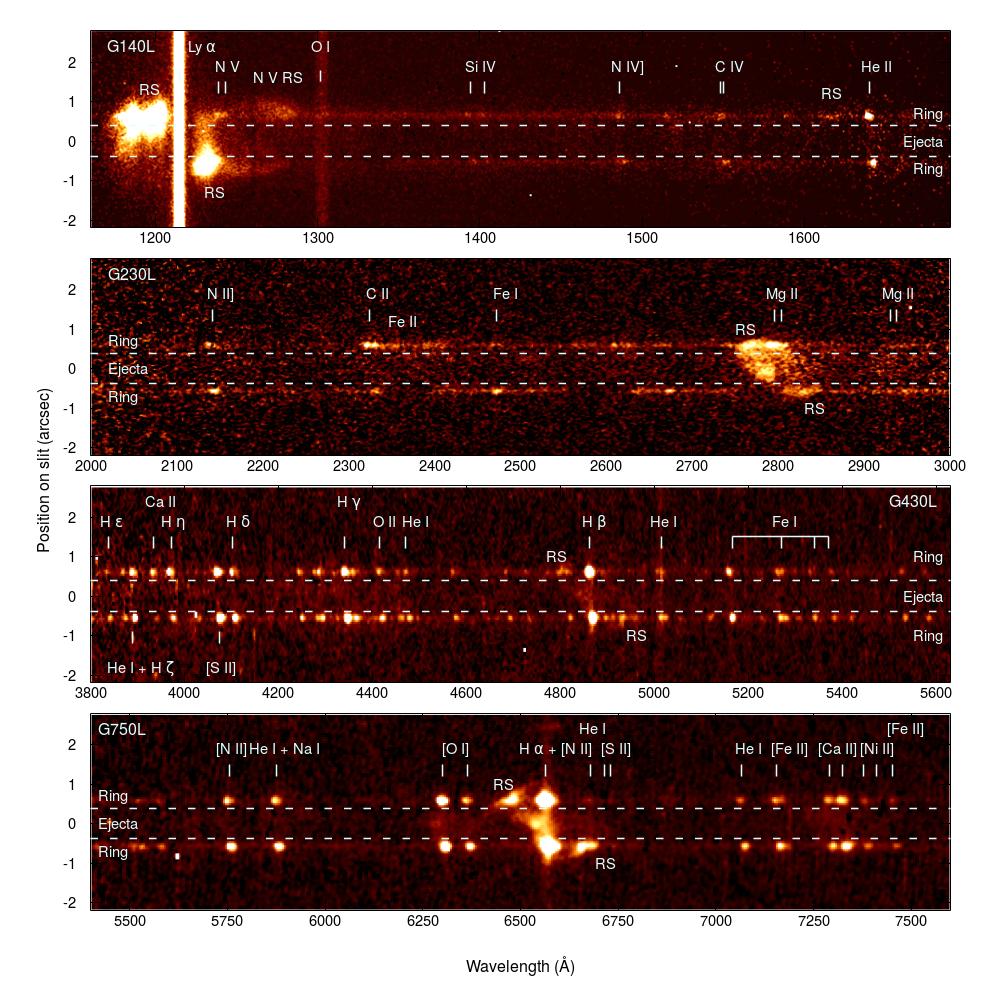}
\caption{Our 2D STIS spectra of the ejecta and ER of SN~1987A with identifications of emission lines. The dashed lines show the extent of the extraction window of our ejecta spectra. The G140L spectrum was taken at Position 2, while the spectra with other gratings are from Position 3.}
\label{fig:pos3_2dspec}
\end{minipage}
\end{figure*}

We used the Python-based {\sc stistools}\footnote{https://github.com/spacetelescope/stistools} package to reduce the STIS spectra. We corrected the spectra for cosmic rays using the \texttt{ocrreject} task, then used the \texttt{x2d} task to obtain wavelength- and flux-calibrated 2D spectra. The 2D spectra were median-combined and 1D spectra were extracted from these using a Python script based on {\sc astropy}\footnote{https://www.astropy.org/}. A background spectrum, extracted from empty regions to the north and south of the ORs, was subtracted from all extractions. As the spectra in Position 2 only constitute two exposures, a median combination was not possible. Instead, we used the \texttt{fixpix} task in {\sc iraf}\footnote{{\sc iraf} is distributed by the National Optical Astronomy Observatory, which is operated by the Association of Universities for Research in Astronomy (AURA) under cooperative agreement with the National Science Foundation.} to remove bad pixels in the G430L and G750L \texttt{x2d} spectra at Position 2, which were then averaged. The reduced STIS 2D spectra, covering the entire wavelength range from 1150 to 10000~\AA, are presented in Fig.~\ref{fig:pos3_2dspec}. In this figure we show the slit at Position 3 except for the G140L spectrum (only taken at Position 2), as this is the spectrum with the longest exposure and the best quality. We also show our extraction window for the ejecta, which we further divide evenly into four bins in the north-south direction in our analysis.

Following the observing plan detailed in~\citet{france11}, the COS position angle was chosen to maximize the spatial separation on the COS detector between Star 3, a Be star bright in the far ultraviolet (FUV), and the main ejecta, RS, and ER regions of the 1987A SNR. Light from objects outside the nominal 1\farcs25 COS aperture radius can be recorded with the science spectrum in crowded fields; the position angle was chosen to prevent significant spatial overlap by placing Star 3 at the ``bottom'' of the microchannel plate (MCP) detector and a custom ``y-walk'' correction was implemented to maximize the angular resolution performance of COS and enable the clean separation of SN 1987A and Star 3.  The COS position angle for all G130M and G160M observations was $\approx-59^\circ$, midway between the $\approx-75^\circ$ and $\approx-45^\circ$ used in the 2011 COS observations. Individual spectral extractions of each object were performed using a custom {\sc idl} script, and the clean spatial separation was verified by comparing the spectra of SN1987A and Star 3; the only contamination observed was a mild overlap of broad Ly$\alpha$ emission in the extracted Star 3 region.

\begin{figure*}
\begin{minipage}{\linewidth}
\centering
\includegraphics[width=\columnwidth]{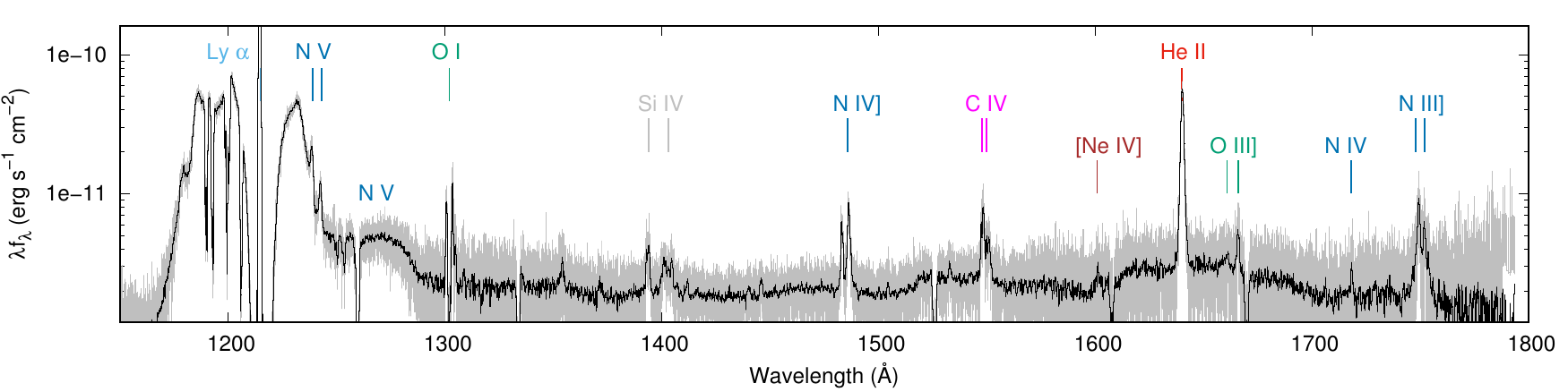}
\includegraphics[width=\columnwidth]{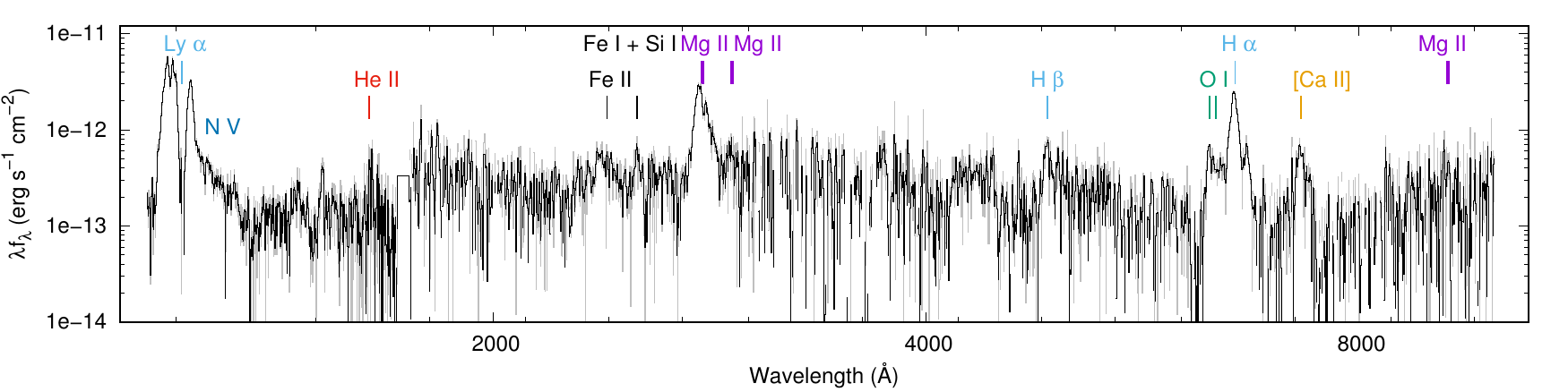}
\caption{Spectra from COS (upper panel; only FUV) and STIS (lower panel; only inner ejecta). ER contribution to the ejecta has been subtracted from the STIS spectrum, while the COS extraction includes the ER. Both spectra have been smoothed for clarity (black) using the Savitzky-Golay algorithm \citep{savgol64}; the original spectrum is also shown (grey). No lines clearly attributable to the inner ejecta are seen in the COS spectrum.  }
\label{fig:total_ejecta}
\end{minipage}
\end{figure*}

The full extracted 1D spectra from both COS and STIS are presented in Fig.~\ref{fig:total_ejecta}. In the STIS spectrum, we have combined Positions 1 and 3, which together correspond to the full extent of the ejecta covered by our observations, while the COS spectrum covers both the ejecta inside the ER and the ER itself. In this and subsequent STIS spectra, we have corrected for the tails of the point spread function of ER light as follows: we found scaling factors for the northern and southern ER separately so that their combined H$\alpha$ flux (where the ER contribution is much more apparent than in e.g. the Mg~{\sc ii}~$\lambda\lambda2796,2804$ doublet) matches the narrow component in each individual ejecta spectrum, then subtracted the ER spectra multiplied by these factors at \textit{all} wavelengths. Due to the brightness of the ER compared to the ejecta, the scaling factors were typically between 0.01 and 0.04, depending on the north-south position. All plotted spectra have been corrected for the recession velocity of SN~1987A, 287~km~s$^{-1}$ \citep[e.g.][]{groning08}. No reddening correction has been applied to the plotted spectra. We note that we refer to all spectral lines by their vacuum wavelength.

\section{Analysis and results}
\label{sec:res}

In this Section, we examine the combined 2018 ejecta spectrum of SN~1987A and compare it to earlier ejecta spectra (Sect.~\ref{sec:spec}). We also construct ejecta velocity maps and compare the morphologies of different line fluxes (Sect.~\ref{sec:velmap}). Finally, we examine the spectrum of the hotspots immediately south of the ER (Sect.~\ref{sec:spots}).

\subsection{Properties and evolution of the full spectrum}
\label{sec:spec}

In the following, when comparing spectra from different epochs, a few caveats should be kept in mind. First, over time, starting from the appearance of the hotspots around the year 2000, an increasing fraction of the ejecta has left the inside of the projected ER and is thus not included in these measurements. The brightest emission has always originated in the inner ejecta, and almost all the emission in a given line is contained within 5000~km~s$^{-1}$. All of the ejecta within this velocity range were still inside the projected ER in 2004. Assuming spherically symmetric expansion, the fraction of the $<5000$~km~s$^{-1}$ ejecta inside the projected ER, in terms of volume, was $\sim0.6$ and $\sim0.4$ in 2014 and 2018, respectively. While the true expansion is not spherical, and the brightness of the ejecta increases toward zero line-of-sight velocity, there is an effect on the fluxes from inside the projected ER at these epochs. The observed brightening of the ejecta (see below) is likely underestimated because of this effect, but as the projection in 2018 still included all ejecta within 3500~km~s$^{-1}$ and thus most of the emission, the effect is still small and we have attempted to correct for it. 

Secondly, based on the slit positions in the F657N image in Fig.~\ref{fig:ha_img} and a F625W image from the same epoch, both from \citet[][]{larsson19}, we estimate that roughly a third of the total ejecta H$\alpha$ flux falls outside the slits, especially from the western bright clump. At other wavelengths, the fraction depends on whether a given line originates in the centre or the edges of the projected unshocked ejecta. We have none the less approximately corrected for this effect by applying a multiplier of 1.5 to the displayed total ejecta spectrum from 2018. Finally, the ER subtraction described in Sect.~\ref{sec:data} may result in over- or undersubtraction of the ER lines, and this effect is strongest in H$\alpha$ where the ER is relatively brightest. An oversubtracted line would appear narrower and fainter than it should be, and vice versa.

\begin{figure*}
\begin{minipage}{\linewidth}
\centering
\includegraphics[width=\columnwidth]{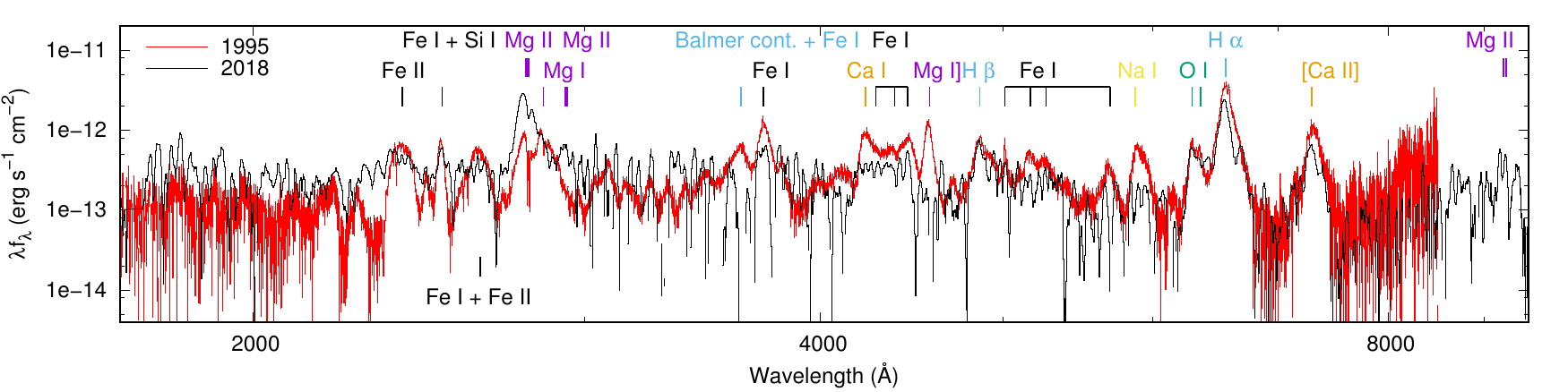}
\caption{Ejecta spectra from the NUV to red wavelengths in 1995 (red) and 2018 (black). The 2018 spectrum has been smoothed for clarity using the Savitzky-Golay algorithm. Line identifications are based on \citet{jerkstrand11}.}
\label{fig:total_ejecta_comp}
\end{minipage}
\end{figure*}

\begin{figure*}
\begin{minipage}{0.99\linewidth}
\centering
\includegraphics[width=0.97\columnwidth]{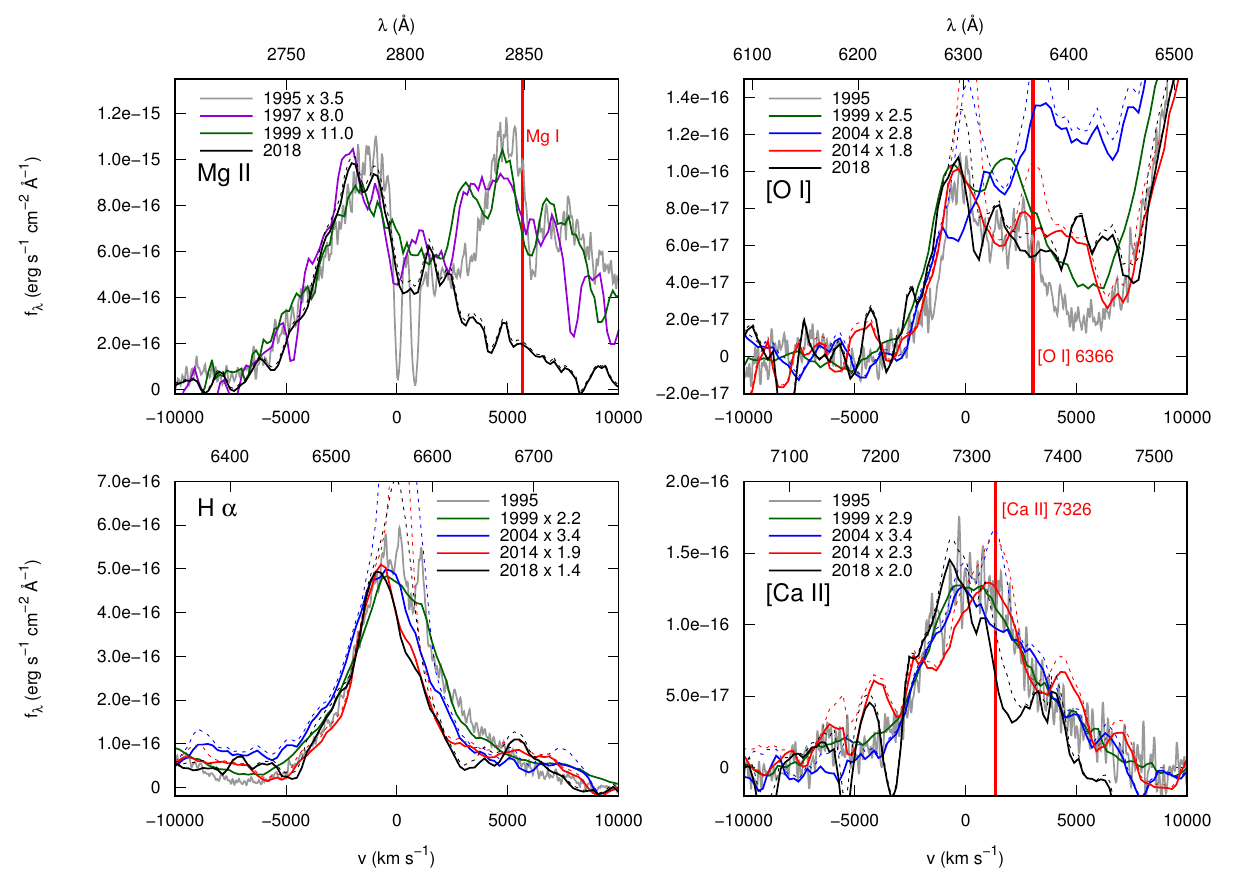}
\caption{Evolution of the strongest ejecta emission lines between 1995 and 2018. The continua (approximated as a separate constant in each panel) have been subtracted. Dashed lines correspond to spectra with no subtraction of ER contamination, while solid lines have been corrected for it. Zero velocity in case of doublets corresponds to the rest wavelength of the bluer line of the doublet. Savitzky-Golay smoothing has been applied to all STIS spectra.}
\label{fig:total_ejecta_lines}
\end{minipage}
\end{figure*}

\begin{figure*}
\begin{minipage}{0.99\linewidth}
\centering
\includegraphics[width=0.97\columnwidth]{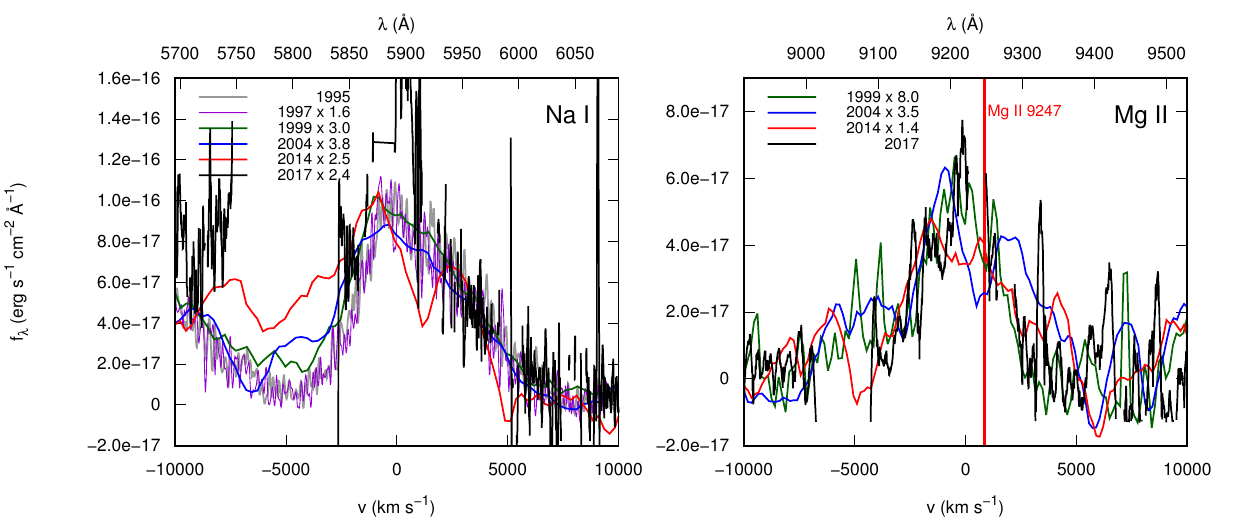}
\caption{Evolution of the Na~{\sc i}~$\lambda\lambda5892,5898$ and Mg~{\sc ii}~$\lambda\lambda9221,9247$ doublets between 1995 and 2017 (neither line is detected in our STIS spectrum). The continua (approximated as constants in each panel) have been subtracted. The evolution of the line fluxes is similar to H$\alpha$ and the Mg~{\sc ii}~$\lambda\lambda2796,2804$ doublet, respectively. The 2017 UVES spectra have been binned and the narrow ER lines removed manually for clarity. Zero velocity corresponds to the rest wavelength of the bluer line of each doublet. Savitzky-Golay smoothing has been applied to all STIS and UVES spectra.}
\label{fig:ejecta_lines2}
\end{minipage}
\end{figure*}

Fig.~\ref{fig:total_ejecta_comp} shows the changes in the spectrum of the ejecta (more precisely, the ejecta still inside the ER) between 1995 \citep{chugai97} -- the last time the wavelength range from 1600 to 8500~\AA~was simultaneously observed -- and 2018. Many of the strong lines from 1995 were below the noise level of the observations in 2018; this includes a host of Fe lines (the lines around 2500~\AA \ are still weakly visible, however), Mg {\sc i} and Na {\sc i}, while only Mg {\sc ii} had brightened. It is clear from our STIS spectrum that the lines in it strong enough to study are Mg~{\sc ii}~$\lambda\lambda2796,2804$, [O~{\sc i}]~$\lambda\lambda6302,6366$, H$\alpha$ and [Ca~{\sc ii}]~$\lambda\lambda7293,7326$. The evolution of these four lines is shown in more detail in Fig.~\ref{fig:total_ejecta_lines}, where we compare the line profiles in 2018 against those in 1995, 1999, 2004 and 2014 \citep[][]{chugai97, heng06, larsson16}. Fig.~\ref{fig:ejecta_lines2}, on the other hand, shows the evolution of the Na~{\sc i}~$\lambda\lambda5892,5898$ and Mg~{\sc ii}~$\lambda\lambda9221,9247$ doublets between 1995 and 2017. These lines are unclear in the 2018 STIS spectrum\footnote{ Na~{\sc i}~$\lambda\lambda5892,5898$ has faded below noise level, while Mg~{\sc ii}~$\lambda\lambda9221,9247$ has brightened over time, but is none the less very noisy and only visible in a smoothed spectrum.}, but visible in our deeper 2017 UVES spectrum \citep{larsson19}.

\begin{figure}
\centering
\includegraphics[width=\columnwidth]{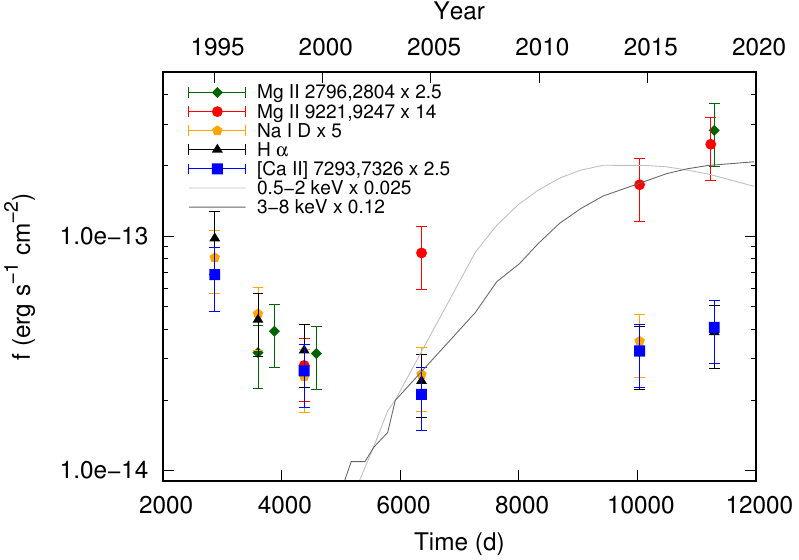}
\caption{Flux evolution of the ejecta emission lines (points), corrected for a combined reddening of $E (B-V) = 0.19$ mag. We have conservatively set the uncertainty at 30 per cent, caused by absolute flux calibration, the removal of ER and RS flux, the changing fraction of ejecta within the projected ER, and other blended components. The fluxes are scaled to facilitate comparisons between their evolution. We also plot soft and hard X-ray fluxes from the ejecta-ER interaction (lines) between 5036 and 5920 d from \citet{frank16} and since 5920~d from \citet{alp21}.}
\label{fig:linefluxes}
\end{figure}

\begin{table*}
\begin{minipage}{0.85\linewidth}
\centering
\begin{small}
\caption{Measured emission line fluxes at each epoch. The fluxes have been corrected for a total reddening of $E(B-V) = 0.19$~mag. All fluxes are in units of $10^{-15}$~erg~s$^{-1}$~cm$^{-2}$. Errors have been estimated at 30 per cent (see Sect.~\ref{sec:spec}).
}
\begin{tabular}{lccccc}
\hline
Epoch (d) & Mg~{\sc ii}~$\lambda \lambda2796,2804$ & Na~{\sc i}~$\lambda \lambda5892,5898$ & H$\alpha$ & [Ca~{\sc ii}]~$\lambda \lambda7293,7326$ & Mg~{\sc ii}~$\lambda \lambda9221,9247$ \\
 \hline
 2875 & $27.4 \pm 8.3$ & $16.2 \pm 4.9$ & $97.7 \pm 29.3$ & $27.3 \pm 8.2$ & - \\
 3604 & $12.7 \pm 3.9$ & $9.3 \pm 2.8$ & $43.8 \pm 13.2$ &- & -\\
 3880 & $15.7 \pm 4.8$ & - & - & - & -\\
 4381 & - & $5.0 \pm 1.6$ & $32.4 \pm 9.8$ & $10.6 \pm 3.2$ & $1.4 \pm 0.5$ \\
 4588 & $12.6 \pm 3.8$ & - & - & - & - \\
 6360 & - & $5.1 \pm 1.6$ & $24.0 \pm 7.3$ & $8.4 \pm 2.6$ & $4.2 \pm 1.3$ \\
 10037 & - & $7.1 \pm 2.2$ & $31.5 \pm 9.5$ & $12.9 \pm 3.9$ & $8.3 \pm 2.5$ \\
 11236 & - & - & -  & - & $12.3 \pm 3.7$ \\
 11295 & $113 \pm 34$ & - & $38.8 \pm 11.7$ & $16.2 \pm 4.9$ & - \\
 \hline
\end{tabular}
\label{tab:lineflux}
\end{small}
\end{minipage}
\end{table*}

The flux evolution of the strongest ejecta emission lines, roughly visible in Figs. \ref{fig:total_ejecta_lines} and \ref{fig:ejecta_lines2}, is shown more clearly in Fig.~\ref{fig:linefluxes}. These fluxes, listed in Table~\ref{tab:lineflux}, have been corrected using a combined Milky Way and LMC extinction of $E(B-V) = 0.19$ mag \citep{fitzpatrick99,france11}. We have set a conservative estimate for the uncertainty of flux measurements at 30 per cent, owing to the approximate subtraction of ER emission and de-blending of nearby lines. We omit [O~{\sc i}]~$\lambda\lambda6302,6366$ due to the blending with H$\alpha$. The fluxes of the H$\alpha$, Na~{\sc i}~$\lambda\lambda5892,5898$ and [Ca~{\sc ii}]~$\lambda\lambda7293,7326$ lines have evolved in a qualitatively consistent way: the flux decreased by a factor of $\sim$3.5 between 1995 and 2004, then gradually increased until 2018 by a factor of $\sim$1.7. While the [O~{\sc i}] flux is more difficult to measure, its brightening is consistent with a factor close to 1.7 as well. The full width half maximum (FWHM, measured relative to the level of the RS emission) of H$\alpha$ decreased over time from $\sim4000$ to $\sim3000$~km~s$^{-1}$ between 2004 and 2014 \citep[mostly due to the brightening of the strong western clump;][]{larsson16}, while in the other lines, no significant FWHM evolution is seen. [Ca~{\sc ii}], in particular, shows practically no change in the shape of the line profile. 

On the other hand, the Mg~{\sc ii}~$\lambda\lambda2796,2804$ doublet flux first decreased by a factor of $\sim2$ between 1995 and 1999, consistently with the other lines, but then increased by a factor of $\sim 9$ between 1999 and 2018. No significant change is seen in the shape of the Mg~{\sc ii} doublet, while the Mg~{\sc i} $\lambda2853$ line stayed strong relative to it at least until 1999 before fading considerably; its contribution was eventually overwhelmed by Mg~{\sc ii} before 2018. Unfortunately this evolution can not be followed between 1999 and 2018\footnote{The \textit{HST} archive does include observations with the G230L grating in 2001, 2002 and 2009; however, the 2001 observation only includes hot spots in the ER, while in the 2002 and 2009 data the ejecta contribution is difficult to disentangle from other components, as the 2\arcsec~slit was used, and any extraction window includes a velocity-shifted contribution by the RS and half of the ER as well.} -- although, as \citet{larsson13} show, UV fluxes around these wavelengths did rise by a factor of $\sim$5 between 2000 and 2009. We can, however, note that the Mg~{\sc ii}~$\lambda\lambda9221,9247$ doublet flux has also brightened over time, roughly by the same factor as the Mg~{\sc ii}~$\lambda\lambda2796,2804$ doublet (i.e. strong brightening by a factor of $\sim$9 since 1999; earlier information is unavailable), as seen in Figs. \ref{fig:ejecta_lines2} and \ref{fig:linefluxes}, and for this doublet we do have measurements in 2004 and 2014 as well. Overall, the flux evolution described here is consistent with what \citet{fransson13} (see their Figs. 8 and 20) and \citet{larsson19} (their Fig.~11) report -- based on line fluxes until 2011 and broad-band photometry until 2018, respectively -- considering the uncertainties and the different extraction regions used.

FUV H$_2$ emission lines are not visible in the COS spectrum (upper panel of Fig.~\ref{fig:total_ejecta}). The presence of these lines would help distinguish between different excitation mechanisms of the near-infrared (NIR) H$_2$ lines reported by \citet{fransson16}. Depending on the excitation mechanism, they should manifest as a series of features between 1200 and 1650 \AA \ \citep[e.g.][]{sternberg89,cmc95,dols00}. The COS spectrum does exhibit emission features, but they are either narrow and identifiable as atomic/ionic lines in the ER, or extremely broad and consistent with RS emission of N~{\sc iv}], C~{\sc iv} and He~{\sc ii}, falling off quickly at blueshifts of $>5000$~km~s$^{-1}$ and forming a pseudo-continuum between 1450 and 1700~\AA. The expected strongest peak of the H$_2$ molecular spectrum at 1608~\AA \ \citep{sternberg89,witt89}, in particular, is not visible -- we do detect an emission line at 1601~\AA, but this is narrow and can be identified as [Ne~{\sc iv}]~$\lambda1601$ instead. On the other hand, an absorption line is visible at 1608~\AA \ (most likely Fe~{\sc ii}~$\lambda1608$), making an upper limit determination for the peak difficult. The 3$\sigma$ upper limit for a possibly expected second peak at 1578~\AA \ \citep[][]{france05}, estimated by summing the error spectrum in quadrature in the window from 1571 to 1584~\AA \ corresponding to a width of 2500~\kms, is $<7.1\times10^{-16}$~erg~s$^{-1}$~cm$^{-2}$.

\begin{figure*}
\begin{minipage}{\linewidth}
\centering
\includegraphics[width=0.95\columnwidth]{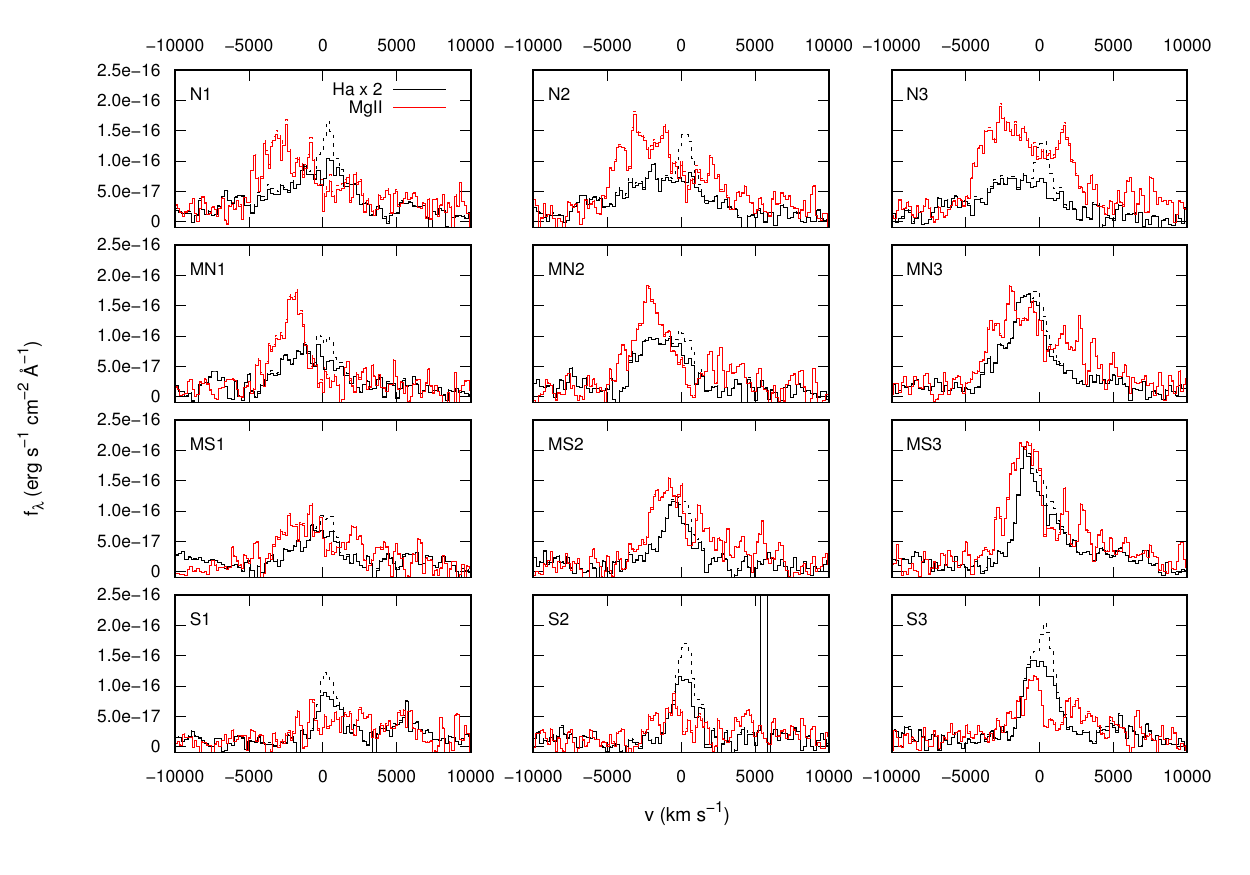}
\caption{Velocity profiles of the H$\alpha$ (black) and Mg~{\sc ii}~$\lambda\lambda2796,2804$ (red) lines in 2018 depending on slit position, as outlined in Fig.\ref{fig:ha_img}, and north-south location. We label the positions from north to south N, MN, MS and S, for 'north', 'mid-north', 'mid-south' and 'south' respectively. Dashed lines correspond to spectra without ER subtraction. A difference in the ejecta structure is visible: the Mg~{\sc ii} line is relatively stronger in the north and north-west, and weaker in the south. Predominantly redshifted emission is only seen in the south-eastern region; the feature at $\sim5000$~km~s$^{-1}$ in the south is attributed to the RS.}
\label{fig:velgrid}
\end{minipage}
\end{figure*}

\subsection{Ejecta morphology and its evolution}
\label{sec:velmap}

In Fig.~\ref{fig:velgrid}, we show the velocity profile of H$\alpha$ and the Mg~{\sc ii}~$\lambda\lambda2796,2804$ doublet in different parts of the ejecta. We separate the full ejecta extraction window (as denoted in Fig.~\ref{fig:pos3_2dspec}) of each slit position into four north-south bins; these are labelled N, MN, MS and S, for 'north', 'mid-north', 'mid-south' and 'south' respectively. Both lines are more centrally peaked in the west and predominantly blueshifted except for the southernmost extraction window, and both show relatively weakest emission in the southeast. Structural differences are none the less visible: the Mg~{\sc ii} peak is sharper and bluer in the east, and Mg~{\sc ii} is relatively stronger toward the north (and northwest) and weaker toward the south compared to H$\alpha$.

In addition to the 1D spectra, we have constructed 'side view' velocity maps at the H$\alpha$, [O~{\sc i}] and Mg~{\sc ii}~$\lambda\lambda2798,2803$ lines to study the structure of the ejecta and compare the different lines. These were made similarly to those in \citet{larsson16}. From the average or median-combined 2D spectra, we subtracted the median background spectrum (extracted north and south of the ORs) from each row and converted the wavelength to a line-of-sight velocity. The distance in the north-south direction from the central point between the northern and southern ER in the 2D spectrum (determined separately for each slit position and line) was converted to north-south velocity assuming homologous expansion. As the ejecta expand freely, the velocity-space morphology of a line directly corresponds to the spatial distribution of the emitting matter. An angular distance of  0\farcs1 corresponds to 0.024 pc in the LMC. At 11300 days, 0\farcs1 in the $y$ direction then translates to $\frac{0.024~\mathrm{pc}}{11300~\mathrm{d}} \approx 760$~km~s$^{-1}$ -- this is the north-south resolution of the velocity map. Along the line of sight the resolution in H$\alpha$ is similar, $\sim$650~km~s$^{-1}$, as measured from the unresolved ER lines in our STIS spectra. Maps at different wavelengths, and using observing setups with different pixel scales, were aligned to the velocities at the H$\alpha$ line using a custom Python script to facilitate comparison and subtraction between maps. 

\begin{figure*}
\begin{minipage}{\linewidth}
\centering
\includegraphics[width=\columnwidth]{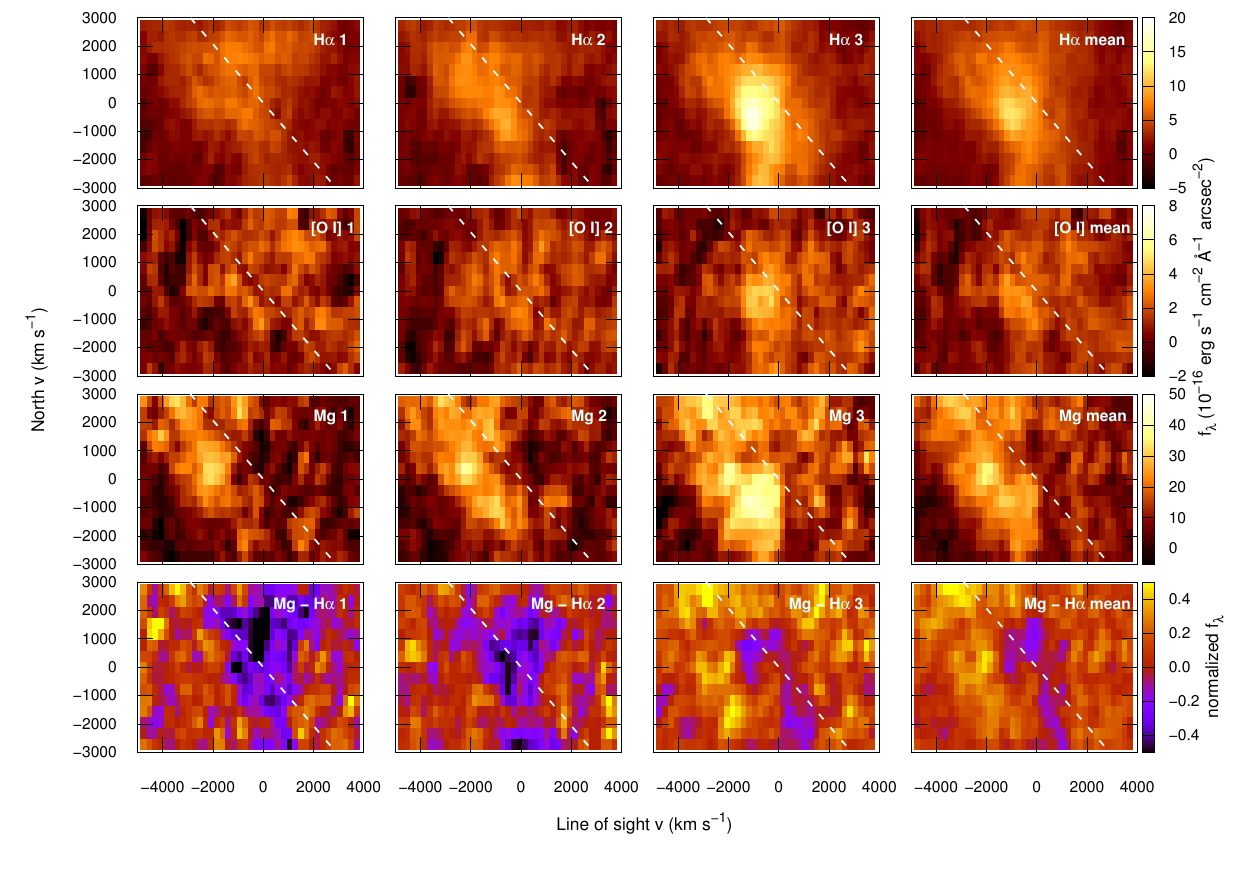}
\caption{Side-view velocity maps of the H$\alpha$ (first row), [O~{\sc i}]~$\lambda6302$ (second row) and Mg~{\sc ii}~$\lambda\lambda2796,2804$ (third row) lines, and a difference map between Mg~{\sc ii} and H$\alpha$ emission (fourth row), at each position and an average of Positions 1 and 3 to represent the total flux. Contribution from [O~{\sc i}]~$\lambda6366$ can be seen in the second row $\sim$3000~km~s$^{-1}$ redward of [O~{\sc i}]~$\lambda6302$. Both Mg~{\sc ii} and H$\alpha$ were normalized to the relative peaks at each position before subtraction. The dashed line indicates the plane of the ER. The observer direction is to the left, and north is up. 1000~km~s$^{-1}$ in the y axis corresponds to roughly $0\farcs13$. All maps correspond to 2018.}
\label{fig:velmapgrid}
\end{minipage}
\end{figure*}

\begin{figure*}
\begin{minipage}{\linewidth}
\centering
\includegraphics[width=\columnwidth]{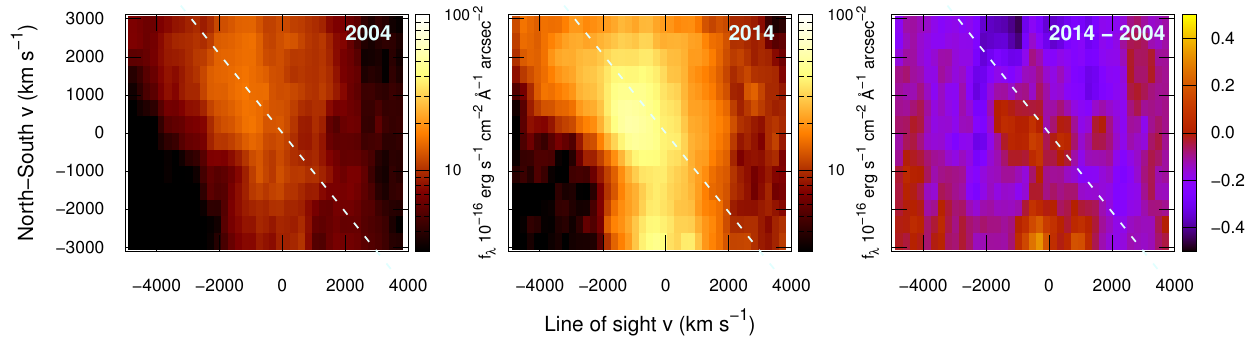}
\caption{Side-view velocity maps of the evolution of the H$\alpha$ line from the inner ejecta, combining all slits in each case. The maps of both years were normalized to their relative peaks before subtraction. The observer direction is to the left, and north is up. }
\label{fig:velmapsub}
\end{minipage}
\end{figure*}

As described in Sect.~\ref{sec:data}, the contribution of the ER in the ejecta spectra was removed by subtracting the northern and southern ER spectra from them, each multiplied by a scaling factor. Similarly, the ER contribution was subtracted from the velocity maps following the example of \citet{larsson16}; the scaling factors were determined for each row of the velocity maps separately and the scaled ER spectrum was subtracted from that row, repeating the process for each row of each H$\alpha$ velocity map. The same scaling factors were used for corresponding rows (i.e. north-south velocities) of the maps at other wavelengths. The final velocity maps are shown in Fig.~\ref{fig:velmapgrid}, including an average of Positions 1 and 3. We also show a subtraction between the Mg~{\sc ii} and H$\alpha$ maps; this was done by normalizing both maps to their peak fluxes before subtraction. 

The velocity maps of H$\alpha$ and [O~{\sc i}] are qualitatively similar, showing an elongated, broken-dipole structure close to the plane of the ER, but blueshifted relative to it by $\sim1000$~km~s$^{-1}$, similarly to what was seen in 2014 \citep{larsson16}. We note, however, that in both lines there is also a redshifted structure visible on the north side of the ER in Position 1. In Position 3 (the western side), we do not see this emission region, but there is a brighter region toward the centre of the ejecta (corresponding to the bright western ejecta in Fig.~\ref{fig:ha_img}), and the structure is slanted toward the plane of the sky in the south, further from the plane of the ER. In Mg~{\sc ii}, the structure is similar in shape, albeit blueshifted toward higher velocities ($\sim2000$~km~s$^{-1}$). We see the same structural differences as in Fig.~\ref{fig:velgrid}, with a relatively stronger Mg~{\sc ii} toward the north and northwest. The direction of stronger Mg~{\sc ii} also matches the strength of the RS and ER emission seen in Mg~{\sc ii}, which are both significantly brighter on the north side -- in fact, the $\sim$zero-velocity component of the Mg~{\sc ii}~$\lambda\lambda2796,2804$ doublet is extremely weak in the southern ER.

As velocity maps let us compare observations from different epochs despite the changing dimensions of the ejecta, they were also constructed in the H$\alpha$ line for the years 2004 \citep{heng06} and 2014 \citep{larsson16} the same way as for 2018. We show these maps in Fig.~\ref{fig:velmapsub}. The velocities in 2004 were aligned to those in 2014 to facilitate comparison. For both years, we have added together all the slit positions of each year, as individual slit positions are not directly comparable. As our observations from 2018 are missing the eastern and western edges of the ejecta, we do not use the 2018 map in this comparison. Changes in morphology between 2014 and 2018 are, however, small, as can be seen by comparing our Fig.~\ref{fig:velmapgrid} to Fig.~10 in \citet{larsson16}. The changes in the structure of the inner ejecta (inside a north-south velocity of 3000~km~s$^{-1}$) between 2004 and 2014 are mostly concentrated in the brightening southern part. The shape is otherwise roughly the same, but the emission brightens over time as described in \citet{larsson19}. Apart from the south, the brightening is relatively strongest in the central part of the side view, corresponding to the dominant western part of the ejecta.

\subsection{Outer spots}
\label{sec:spots}

\begin{figure}
\centering
\includegraphics[width=\columnwidth]{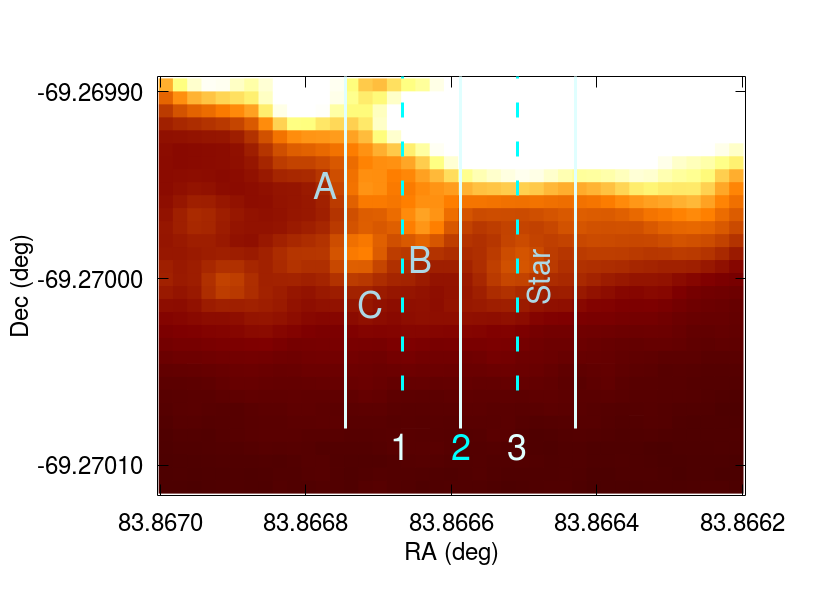}
\caption{Zoomed-in cutout of Fig.~\ref{fig:ha_img} showing three outer hotspots outside the ER covered by our slit Position 1; from north to south, we label them A, B and C. Spot B is also covered in Position 2.}
\label{fig:ha_img_zoom}
\end{figure}

\begin{figure}
\includegraphics[width=0.95\columnwidth]{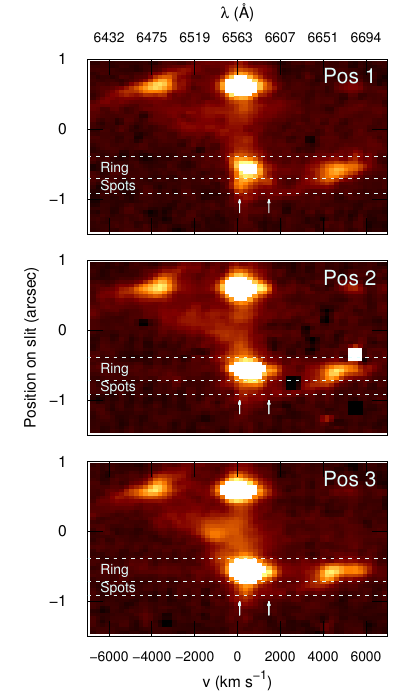}
\caption{Extraction window for our hotspot spectra outside the ER at H$\alpha$, and a comparison between positions. Two emission peaks, at the declinations of two hotspots (B and C), are visible in Position 1, indicated with arrows. In Positions 2 and 3, these features are not replicated.}
\label{fig:spotext}
\end{figure}

\begin{figure}
\centering
\includegraphics[width=\columnwidth]{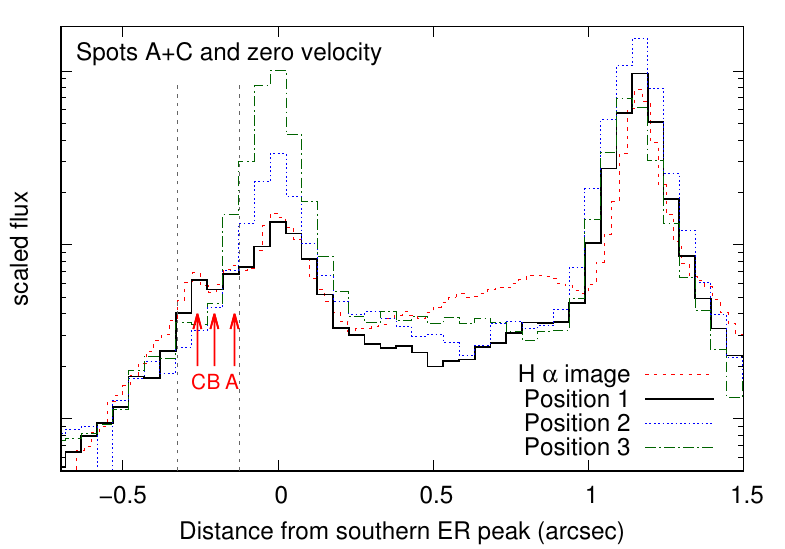}
\includegraphics[width=\columnwidth]{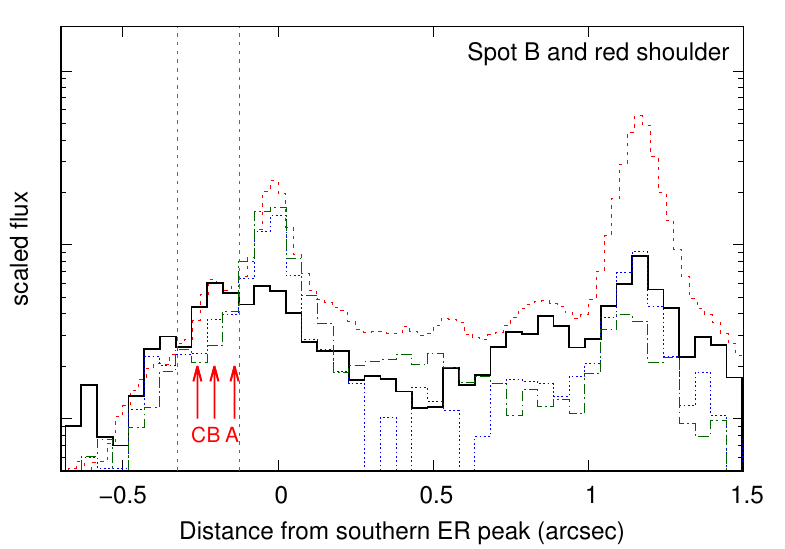}
\caption{Vertical slices of the H$\alpha$ line map (Fig.~\ref{fig:ha_img} and \ref{fig:ha_img_zoom}) and the 2D spectra at H$\alpha$. The slices of the map correspond to the eastern (A and C) and western (B) spots, while the slices of the 2D spectrum are extracted around 0 and 1500~km~s$^{-1}$ respectively (arrows in Fig.~\ref{fig:spotext}). The vertical lines denote the extraction window and the arrows mark positions of spots. Spots B and C correspond to peaks in the vertical 2D spectrum slice in Position 1; such peaks are not seen at other positions, but Position 2 seems to exhibit a shoulder at the position of spot B at 1500~km~s$^{-1}$. Note the logarithmic $y$ axis and arbitrarily scaled fluxes.}
\label{fig:spotdist}
\end{figure}

Three new hotspots outside the ER \citep[seen in imaging observations by][]{fransson15} are covered by our slit in Position 1, as can be seen in Fig.~\ref{fig:ha_img_zoom}. From north to south, we refer to them as spots A, B and C. \citet[][]{larsson19} identified the 'spot' in Position 3 as a star. These spots also seem to be associated with emission visible south of the ER in the Position 1 2D spectrum, and we have extracted their combined spectrum, covering all three spots. We show the extraction window in Fig.~\ref{fig:spotext} at the H$\alpha$ line, where the spots are at their brightest. In the same figure we also compare this to the corresponding spectra at other positions (for comparison, the slit in Position 2 covers only spot B, while the slit in Position 3 covers none). 

A clump corresponding to the declination of spot C is clearly visible in Position 1 in Fig.~\ref{fig:spotext}; spot A is difficult to distinguish from the ER in this 2D spectrum, but is located $\sim0\farcs13$ south of the ER peak in the H$\alpha$ line map (see Fig.~\ref{fig:ha_img_zoom}); this is also covered by the window. The declination of spot B, meanwhile, seems to coincide with a faster clump at $\sim1500$~km~s$^{-1}$ ($\sim6590$~\AA). This clump is also only visible in Position 1. To demonstrate these associations, we show north-south slices at the positions of the hotspots in Fig.~\ref{fig:spotdist}. The eastern slices (top panel) are extracted over the two eastern hotspots (A and C) in the line map and 6555--6575~\AA~($-370$ to $+550$~km~s$^{-1}$) in the 2D spectrum, while the western slice (lower panel) covers the position of spot B and 6585--6605~\AA~(+1000 to +1900~km~s$^{-1}$). The declinations of the two clumps in the 2D spectrum at Position 1 match the locations of spots B and C in the H$\alpha$ image. The 6-month time difference between the STIS spectrum and the H$\alpha$ line map is not expected to be significant here, as these three spots are visible at similar flux levels since roughly 2014 \citep{larsson19}.

\begin{figure*}
\begin{minipage}{\linewidth}
\centering
\includegraphics[width=\columnwidth]{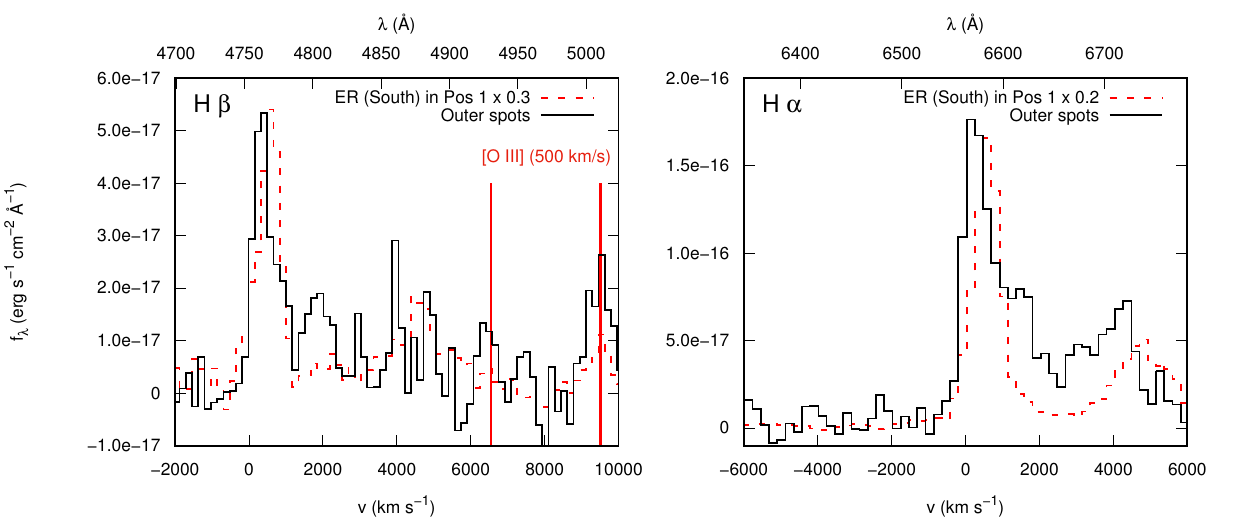}
\caption{Velocity profiles of the H$\beta$ (left) and H$\alpha$ (right) lines at the location of the outer hotspots in Position 1, compared to the southern ER line profiles at the same position. The peak in the narrow component of the line is slightly offset compared to the ER. The velocity interval in the left panel is chosen to show the [O~{\sc iii}] $\lambda\lambda4960,5008$ doublet as well; only the the 5008~\AA~component is visible. The peak velocity of this feature is $\sim$500~km~s$^{-1}$ (red vertical lines); the 5008~\AA~line is relatively stronger compared to H$\beta$ in the spot spectrum, indicating a contribution from the spots. The features at 4000-6000~km~s$^{-1}$ (in H$\beta$, only seen in the ER spectrum) are due to the RS.} 
\label{fig:spotspec}
\end{minipage}
\end{figure*}

The only clearly visible lines in the spectra of the spots, extracted as shown in Fig.~\ref{fig:spotext}, are H$\alpha$ and H$\beta$. We show these line profiles in Fig.~\ref{fig:spotspec} -- we also show wavelengths redward of H$\beta$ including the wavelengths of the [O~{\sc iii}] doublet. The spectra show an unresolved narrow ($\leq670$~km~s$^{-1}$) component peaking at $\sim100$~km~s$^{-1}$, a few hundred~km~s$^{-1}$ blueward from the peak of the ER line and thus unlikely to be dominated by ER emission. A shoulder or bump is also visible on the red side of the narrow component, peaking at $\lesssim2000$~km~s$^{-1}$ in both lines, but the noisiness at H$\beta$ makes this uncertain\footnote{The shoulder in H$\alpha$ is located at 6593~\AA~in the LMC rest frame, and thus unlikely to be [N~{\sc ii}]~$\lambda6584$ from the same spots even if the noisy H$\beta$ shoulder is coincidental.}. In addition, we see the influence of the RS peaking at $\sim4000$~km~s$^{-1}$ in H$\alpha$; in H$\beta$ the noise makes the RS peak unclear. A feature can be attributed to [O~{\sc iii}] emission at $\sim500$~km~s$^{-1}$, consistent with the ER velocity; however, the spots likely contribute to this feature as well, as the [O~{\sc iii}] is relatively stronger compared to H$\beta$ in the spots than in the ER as can be seen from Fig.~\ref{fig:spotspec}. The ER extraction includes RS emission from higher velocities, and in the ER it thus peaks at $\sim4500$~km~s$^{-1}$, consistently with the "horn" structure of the RS in recent spectra \citep[e.g.][Fransson et al. in preparation]{larsson19}. 

The continuum level is low after sky subtraction, and the total flux within the \emph{HST} F625W filter is dominated by the H$\alpha$ line as has been suggested for other new spots. The H$\alpha$ line flux in the unresolved narrow component from all three spots, corrected for extinction, is $\sim5\times10^{-15}$~erg~s$^{-1}$~cm$^{-2}$, roughly twice the flux of spot C measured by \citet{larsson19} -- with the caveat that any ER contribution has not been subtracted. The matching declinations between features in the Position 1 2D spectrum and the spots, the shift in the line peak compared to the ER, and the reasonable correspondence between photometric and spectroscopic line fluxes lead us to conclude that we have indeed detected these spots in our spectroscopy. 

The red shoulder, meanwhile, corresponds to a relatively faint peak in the 2D spectrum, the location of which coincides with the declination of the western spot B. However, in Position 2, the higher-velocity feature is not clearly seen, while the slit should cover spot B. A tentative bump is seen at the corresponding declination in the north-south slice of the 2D spectrum, but does not reach the same brightness -- if the faster clump is indeed associated with spot B, one would expect it to be equally bright in both spectra. Thus we conclude that the association between spot B and the higher-velocity shoulder is coincidental. In addition, as seen in Fig.~\ref{fig:spotext}, there is a fainter but visible clump on the northern side, at roughly the same velocity, as well. This clump can also be seen in Fig.~\ref{fig:spotdist}. Neither clump is discernible in the 2D spectra at other positions.

\section{Discussion}
\label{sec:disco}

\subsection{X-ray deposition and cooling of the ejecta}
\label{sec:xray_dep}

It has been established in previous studies of the SN~1987A ejecta that the edge-brightened morphology of the inner, freely expanding ejecta in H$\alpha$ (see e.g. Figs.~\ref{fig:sketch} and ~\ref{fig:ha_img}) is caused by their excitation being dominated by X-rays from the shock of the ER-ejecta interaction that get absorbed outside the $^{44}$Ti-powered core; that lines such as the infrared [Fe~{\sc ii}]+[Si~{\sc i}] at 1.644 $\mu$m, which peak at the centre of the ejecta, are powered by $^{44}$Ti decay; and that dust is unlikely to explain the differences between these structures \citep[][]{kjaer10,larsson11,fransson13,larsson13,larsson16}. Our new observations are consistent with this picture, and suggest the X-ray power source for several other lines as well. From Fig.~\ref{fig:velmapgrid} we see that the distribution of [O~{\sc i}] is fainter but very similar in shape to that of H$\alpha$, while that of Mg~{\sc ii} is slightly blueshifted compared to them. The flux evolution of the lines shown in Fig.~\ref{fig:linefluxes} was qualitatively similar until 1999. After that, the Mg~{\sc ii} lines brightened more than the rest, but all lines evolved in a similar manner from 2004 to 2018. Hard X-rays may be the main power source, as their flux has continued to increase \citep{frank16,alp21}, while the flux from \textit{soft} X-rays reached a peak around the year 2014, followed by a modest decline. 

The observed FWHM of the H$\alpha$ line decreases over time, despite the fact that the FWHM should broaden as ejecta at preferentially low line-of-sight velocities have left the ER or are missing from our slit coverage. This may be in part due to the X-rays penetrating deeper layers of the ejecta over time as the optical depth decreases \citep[][]{fransson13}. The western clump visible as a bright low-velocity feature in Position 3 (Fig.~\ref{fig:velgrid}) may be the dominant factor, though. We do note that any measured FWHM can be affected by over- or undersubtraction of the narrow ER features, especially when it comes to H$\alpha$, where the narrow lines are relatively strong. Additionally, dust distributed among the ejecta preferentially absorbs emission from the far side of the ejecta. If its optical depth decreases over time, this should then result in a slow strengthening of the red wing at each line. We do not observe such evolution (Fig.~\ref{fig:total_ejecta_lines}).

X-ray deposition in the ejecta was discussed in \cite{fransson13}, based on the 14E1 model in \cite{blinnikov00}. We update this with the M15-7b merger model from \cite{utrobin21}, which gives the best agreement with the light curve of SN~1987A. This model is based on a 15 + 7 M$_\odot$ merger model from \cite{menonheger17}, which was exploded in 3D and then mapped into a 1D model by taking spherical averages of the density and abundances. As in \cite{fransson13}, we assume spherical symmetry and a RS located at 80\%\footnote{This is only important to set the maximum radius and velocity for the deposition and emission. The ejecta outside this radius are shocked and will not emit in the optical band.} of the radius of the ER. X-ray photons are assumed to be emitted at the ER. There is also some emission at high latitudes and thus larger distances from the centre, resulting in a larger dilution factor and a lower fraction of hard X-rays being absorbed, but this component is likely to be relatively faint \citep{alp21}. As 
we are mainly interested in the qualitative nature of the effects of the X-rays, we ignore the dilution effects.

\begin{figure*}
\includegraphics[width=1.03\columnwidth]{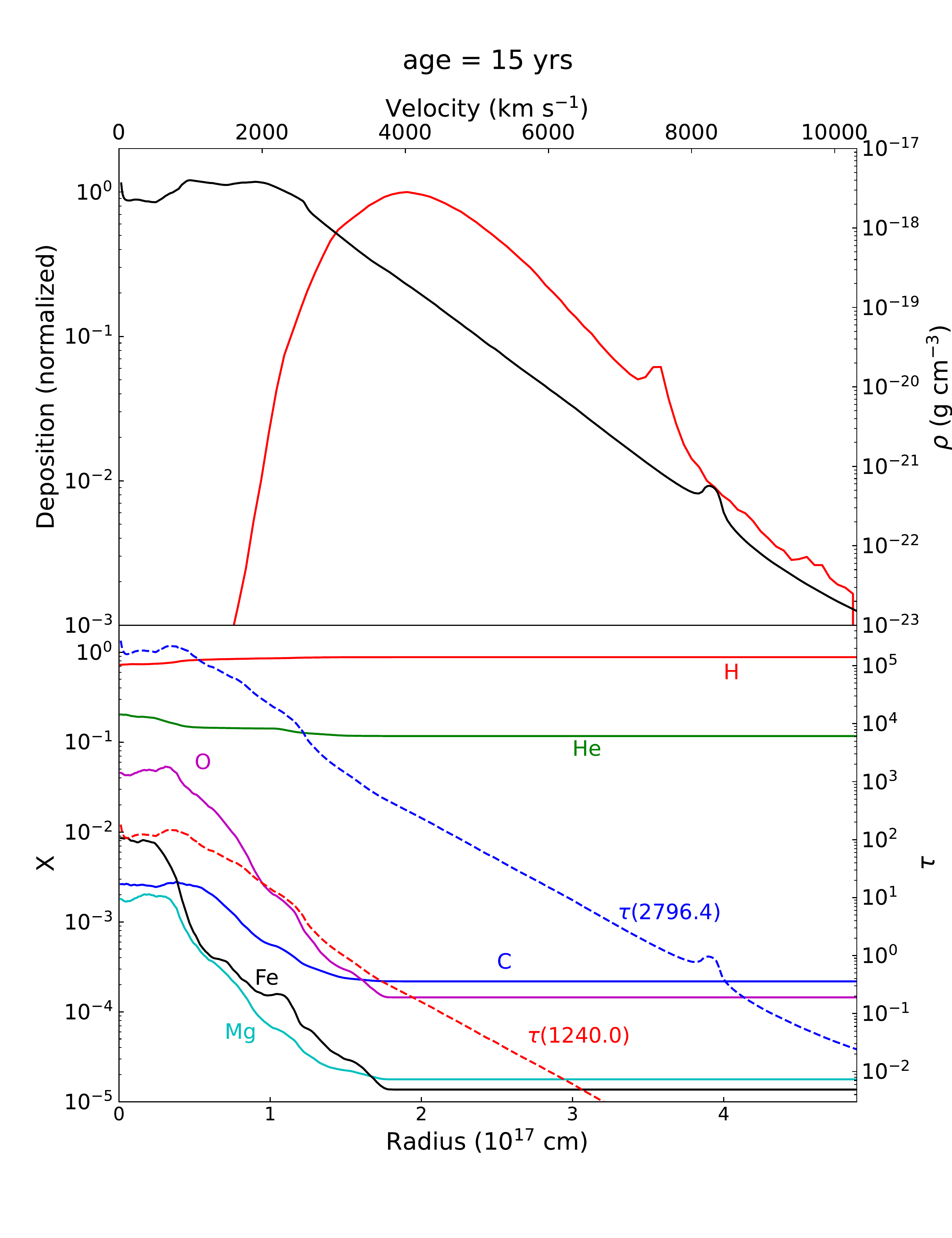}
\includegraphics[width=1.03\columnwidth]{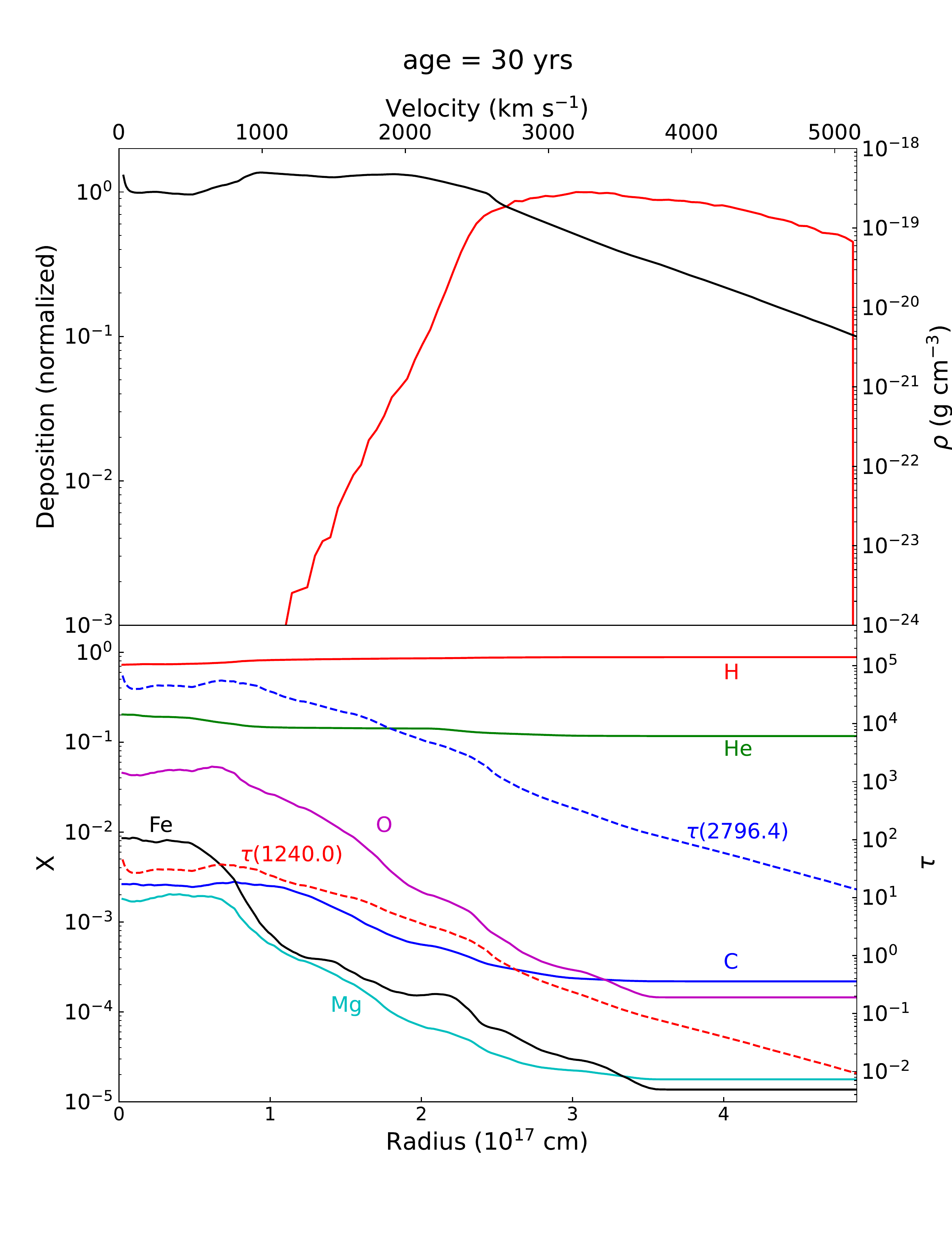}
\caption{Upper panels: Density (black) and the X-ray energy deposition (red) for the M15-7b model from \citet{utrobin21} at 15 yrs and 30 years. Lower panels: Elemental fractions by number for the same model. The dashed red line shows the optical depths in the Mg~{\sc ii} $\protect\lambda 1240.0$ line, responsible for the Ly$\alpha$ fluorescence, while the dashed, blue line shows the optical depth of the  Mg~{\sc ii}$\protect\lambda 2796.4$ line. }
\label{fig:endep}
\end{figure*}

In Fig.~\ref{fig:endep} we show the density and energy deposition at 15 and 30 years and the abundance structure of the ejecta. While \citet{fransson13} showed the deposition only at a few different energies, we show the total deposition from a realistic X-ray spectrum: we take the spectral fit at 12141 days by \cite{alp21}. Although this is slightly more recent than our HST observations at $\sim11000$ days, the spectral shape was shown to change very slowly. For the 15 year model we use the same spectrum because we are mainly interested in the qualitative differences between these epochs, although the absolute flux changes.

From Fig.~\ref{fig:endep} we can more directly compare the effects of the X-ray input to the observed distribution of especially the Mg~{\sc ii} emission as seen in e.g. Figs. \ref{fig:total_ejecta_lines} and \ref{fig:velmapgrid}. Qualitatively, the main difference in energy deposition compared to the 14E1 model in \cite{fransson13} is the use of a continuous spectrum for the deposition. The density profile is slightly steeper at high velocities, flatter in the core, and similar in the range of 2500 - 5000~\kms. There is also more extensive mixing in the M175b model compared to the 14E1 model. Most importantly, X-rays with energy $\ga 1$ keV can penetrate the ejecta to velocities down to $\sim 2000$~\kms, and at somewhat lower velocities the high metal abundance stops X-rays with energies $\la 5$ keV. Recent {\it NuSTAR} observations \citep{alp21} show that the X-rays from the interaction of the ejecta with the CSM extend to at least $\sim 30$ keV. The most obvious difference between the models at 15 and 30 years is that the expansion and the steep density profile of the ejecta have caused the density close to the RS to increase dramatically. At 15 years the RS is at $\sim 10,300$~\kms, while at 30 years it is at $5150$~\kms. In addition, the core is much closer to the ring and a larger fraction of the X-ray flux will therefore be absorbed by the region of high metallicity. This contributes to increasing the flux of the main cooling lines, in particular the Mg~{\sc ii} $\lambda\lambda2796,2804$ lines. 

In spherical symmetry, the low deposition inside the core leads to flat-topped line profiles at velocities $\la 1000-2000$~\kms \ for lines dominated by X-ray input. However, the presence of such a profile at Mg~{\sc ii} $\lambda\lambda2796,2804$ (or other lines) is uncertain due to the limited S/N, and the red side below $\sim 2000$~\kms \ is affected by ISM absorption as seen in Fig.~\ref{fig:total_ejecta_lines}. In reality, clumping also likely leads to deposition at even lower velocities, which may result in a more peaked profile. In relation to this profile, we note that this strong resonance line may be affected by radiative damping in the same way as the Ly $\alpha$ line\footnote{ The damping parameter for the Mg~{\sc ii} $\lambda\lambda2796,2804$ transitions is  $a=2.2 \times 10^{-3}$. The damping wings become important when $\tau a \ga 1$. From Fig.~\ref{fig:endep}, we see that this is the case inside $\sim 2500$~\kms, where $\tau(2796.4) \sim 5 \times 10^{4}$. Clumping may result in even stronger damping within the clumps. In Fig.~\ref{fig:pos3_2dspec} there is an indication of a similar suppression of the red wing. A more quantitative analysis of the effect is outside the scope of this paper.}. This will lead to a broader absorption profile and may suppress the red wing of the line \citep[see e.g.,][for a discussion in relation to Ly$\alpha$]{france15}.

Because of the high Mg abundance in the model (Fig.~\ref{fig:endep}), the likely dominance of singly ionized Mg and low excitation potential of the Mg~{\sc ii}~$\lambda\lambda2796,2804$ lines, these are also expected to be among the most important lines for the cooling of the ejecta. This is supported by the general models of circumstellar interaction in \cite{chevalier94,chevalier17}, where these lines dominate the cooling of the partially ionized regions in shocked ejecta. They have been observed throughout the evolution of interacting SNe \citep{fransson05,fransson14,fox20}, albeit not as late or with the spatial resolution as is doable with SN 1987A. This doublet is indeed the strongest feature in our UV spectrum. Besides the Mg~{\sc ii} lines, UV lines from Fe~{\sc ii} are contributing to the cooling. These are all permitted lines with a low excitation energy, from the dominant ionization stage of an element with relatively high abundance. Because the emission from these is distributed over a large number of transitions between $\sim 2400 - 2700$ \AA \ they do not appear as prominent as the Mg~{\sc ii} lines in the spectrum (Fig.~\ref{fig:total_ejecta_comp}). The H $\alpha$ line is mainly a result of recombination, following X-ray ionizations, and is less important for the cooling.

\subsection{Ly $\alpha$ fluorescence of Mg~{\sc ii}}
\label{sec:fluor}

Understanding the contributions of different power sources to the emission lines carries implications for the abundances in the ejecta, in turn affecting models of nucleosynthesis in SNe such as that of \citet{jerkstrand11}, at all stages of the evolution. With that in mind, we examine the evolution of the lines. We have detected a major brightening in the flux of Mg~{\sc ii}~$\lambda\lambda2796,2804$ compared to H$\alpha$ in our \emph{HST} spectrum, raising the question of what powers this increase. \citet{fransson13} identified the emission feature at $\sim$9230~\AA~as being caused by the Mg~{\sc ii}~$\lambda\lambda9221,9247$ doublet \citep[an identification supported by the synthetic spectrum in][]{jerkstrand11}. They suggested this feature is powered by fluorescence by Ly$\alpha$ or Ly$\beta$ photons, but Ly$\beta$ was disfavored, as it should be accompanied by emission at Mg~{\sc ii}~$\lambda\lambda7879,7898$ or $\lambda\lambda8216,8237$. The absence of these lines also argues against recombination dominating the excitation. 

We find a similar flux evolution between the Mg~{\sc ii}~$\lambda\lambda9221,9247$ and $\lambda\lambda2796,2804$ doublets, suggesting a common excitation mechanism. Until 1999, the decline of Mg~{\sc ii}~$\lambda\lambda2796,2804$ and $\lambda\lambda9221,9247$ is still consistent with that of H$\alpha$. However, from 1999 to late 2017 or early 2018, both doublets increased in flux by a factor of $\sim9$, while the H$\alpha$ fluxes are consistent within the uncertainties at these times, as seen in Fig.~\ref{fig:linefluxes}. 
The Mg~{\sc ii}~$\lambda\lambda9221,9247$ doublet brightened monotonically between 1999 and 2017, instead of declining until 2004, and did so by roughly the same factor as Mg~{\sc ii}~$\lambda\lambda2796,2804$, suggesting that the latter may also have evolved in a similar way. On the other hand, the brightening of the Na~{\sc i}~$\lambda\lambda5892,5898$, [O~{\sc i}]~$\lambda\lambda6302,6366$ and [Ca~{\sc ii}]~$\lambda\lambda7293,7326$ doublets between 2004 and 2018 was similar to that of H$\alpha$. This suggests that these lines and the Mg~{\sc ii} lines are excited by different mechanisms.

The morphology of Mg~{\sc ii}~$\lambda\lambda9221,9247$ was found similar to H$\alpha$ by \citet{larsson16}, although the data were quite noisy -- and the Ly$\alpha$ fluorescence power source should indeed result in a morphology similar to H$\alpha$ \citep{fransson13}\footnote{Ejecta Ly$\alpha$ emission itself, unlike Ly$\alpha$ from the RS, is largely drowned out by interstellar absorption, although the most blueshifted component is due to resonant scattering of photons from the ER by the ejecta \citep{france10}.}. The absorbed Ly$\alpha$ photons should be redshifted, likely originating on the far side of the ejecta and absorbed on the near side, resulting in a preferentially blueshifted distribution. In Mg~{\sc ii}~$\lambda\lambda2796,2804$, the morphology of the emission in Fig.~\ref{fig:velmapgrid} is indeed broadly similar to H$\alpha$, but relatively more blueshifted -- but suppression of the red wing by radiative damping can also contribute to this. Additionally, from Fig.~\ref{fig:velgrid} we can tell that while both Mg~{\sc ii} and H$\alpha$ are generally stronger in the western part of the ejecta, where a strong peak is seen in H$\alpha$ as shown in Fig.~\ref{fig:ha_img}, Mg~{\sc ii} is also relatively stronger than H$\alpha$ in the north and northeast.

To test the fluorescence scenario we have calculated the relative fluxes of the expected Mg~{\sc ii} lines from a model atom including all 15 levels up to the 5p $^2$P level. Radiative transition rates and collision strengths are taken from the CHIANTI atomic database \citep{dere97,dere19}. Optical depths are calculated in the Sobolev approximation. The Ly$\alpha$ fluorescence process is simulated by artificially populating the upper levels with the Mg~{\sc ii}~$\lambda\lambda 1239.9, 1240.4$ doublet (3s $^2$S -  4p $^2$P), and then calculating the relative fluxes of the resulting lines. We assume a temperature of $10000$ K. The saturated ratios of lines relative to the $\lambda 2804$ line flux, and of multiplets to the $\lambda 2798 $ multiplet, are listed in Table~\ref{tab:mgii_fl}. 

\begin{figure}
\includegraphics[width=0.95\columnwidth]{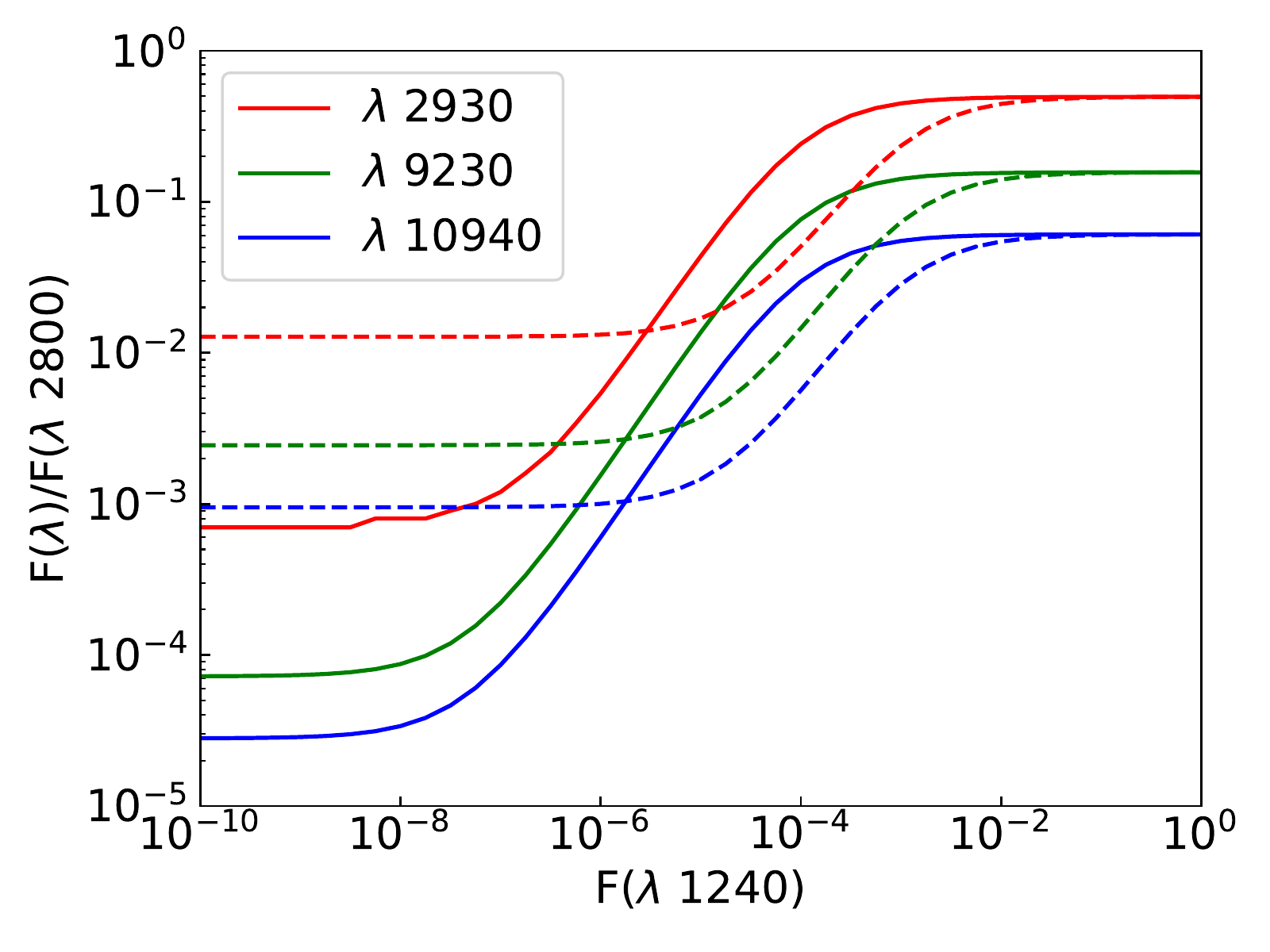}
\caption{Ratios between Mg~{\sc ii} emission line fluxes relative to the $\lambda 2800$ blend as a function of the absorbed $\lambda\lambda 1239.9, 1240.4$ flux. The solid lines are for 10,000 K and the dashed for 20,000 K. The flux units are arbitrary.}
\label{fig:mgiirratios}
\end{figure}

The Doppler broadening in the ejecta is $\ga 2000$~\kms \ and neighboring lines will be blended into a few broad lines, as seen from the spectra. In Fig.~\ref{fig:mgiirratios}  we give the ratios of the most relevant line features, adding the different multiplets due to the blending, relative to the $\sim$2798~\AA~blend for the temperature above. We can see that thermal excitation by collisions -- i.e. in the small $F(\lambda 1240$) case -- is far from sufficient to populate any other levels than the first excited state. As the $\lambda\lambda 1239.9, 1240.4$ flux increases, the ratio relative to the $\lambda\lambda 2796, 2804$ lines saturates when the fluorescence cascade dominates over the thermal excitation of the first excited level. Here we have also calculated line ratios for a temperature of 20000 K to show the maximum thermal excitation of the lines, but the temperature in the ejecta is likely to be considerably lower. When fluorescence is important, the temperature only marginally affects the ratios. 

Comparing to the observations, we note that the Mg~{\sc ii}~$\lambda 1240$ lines are swamped by the Ly $\alpha$ and N~{\sc v} lines from the ring and RS. Furthermore, the spectrum does not cover wavelengths past 10000~\AA, so measuring the flux at $\lambda\lambda10917,10918$ or $\lambda10955$ is not possible. The remaining lines are therefore the $\lambda 2798$, $\lambda\lambda 2930,2937$, and $\lambda\lambda9221,9247$ multiplets. A weak feature in the STIS spectrum can be identified as $\lambda\lambda2930,2937$, peaking at 2931~\AA~with $0.14\pm0.01$ times the flux of the combined $\sim$2800 \AA~feature. Corrected for the Galactic and LMC reddening of $E (B-V) = 0.19$ mag, the flux of the $\lambda\lambda9221,9247$ doublet is $0.07\pm0.02$\footnote{This includes an uncertainty of $\sim10$ per cent in the missing ejecta correction in the 2018 spectrum, which is not required for the $\lambda\lambda9221,9247$ measurement based on the 2017 spectrum.} times the $\sim$2800~\AA~feature. In the saturation scenario, as seen from Table~\ref{tab:mgii_fl}, the predicted flux ratios are 0.49 and 0.16, respectively. Clearly this is not the case based on the measured ratios, suggesting a contribution from another source.

There are indications that the Mg~{\sc ii}~$\lambda\lambda2796,2804$ doublet also has an important thermal contribution, resulting in lower line ratios relative to the $\sim2800$~\AA \ blend (see Fig.~\ref{fig:mgiirratios}). Most importantly, the $\sim2800$~\AA \ blend is broader and slightly more blueshifted than the $\lambda\lambda9221,9247$ doublet (Figs. \ref{fig:total_ejecta_lines} and \ref{fig:ejecta_lines2}), 
extending to $\sim 8000$~\kms \ on the blue side. There is a relatively minor contribution by the Mg~{\sc ii}~$\lambda\lambda2792,2799$ lines, but 
the contribution by the RS or thermal excitation in the ejecta should be dominant. The former, however, is expected to result in a boxy line profile, as seen in e.g., the N~{\sc v} $\lambda 1239,1243$ or  He~{\sc ii} $\lambda 1640$ lines in Fig.~\ref{fig:total_ejecta}, while the  Mg~{\sc ii}~$\lambda\lambda2792,2799$ profile is more peaked, showing that most of the emission is coming from the inner ejecta (Fig.~\ref{fig:total_ejecta_lines}). The latter option is a result of the X-ray flux from the ejecta-ER interaction (Sect.~\ref{sec:xray_dep}), which is also powering the H$\alpha$ line. This is also supported by the roughly similar spatial distributions of the lines seen in Fig.~\ref{fig:velmapgrid}. We therefore believe that thermal excitation is the dominant contribution to the Mg~{\sc ii}~$\lambda\lambda2796,2804$ lines. There should also be a fluorescence contribution, especially from the inner regions where the Mg~{\sc ii}~$\lambda1240$ lines are optically thick -- the measured line ratios are consistent with this amounting to roughly one third of the total excitation. Thus the importance of thermal excitation has unsurprisingly increased since the spectrum at 8 years, before heating by X-rays from the interaction became important and the ejecta temperature was $\lesssim200$~K, where \citet{jerkstrand11} found the feature at $\sim2800$~\AA \ to be dominated by fluorescence. At the same time, the fluorescence changes from being dominated by Ly$\alpha$ from the core region to Ly$\alpha$ and N~{\sc v} from the ER. The line fluxes from fluorescence and thermal excitation, both powered by the ejecta-ER collision, evolve in a qualitatively similar way (Fig. \ref{fig:linefluxes}).

\begin{table}
\centering
\caption{Ratios of Mg~{\sc ii} emission line fluxes when Ly$\alpha$ fluorescence dominates the excitation.
The last row in each multiplet gives the weighted wavelength and total flux relative to the $\lambda 2797.9$ multiplet.}
\begin{tabular}{ccc}
\hline
$\lambda$ (\AA) & $ F(\lambda) / F(\lambda 2803.5$)& $ F(\lambda) / F(\lambda 2797.9$)\\
 \hline
      1239.9&       0.031&\\ 
      1240.4&       0.059&\\
      1240.2&  &     0.025\\       
      &\\
      2791.6&       0.390&\\
      2796.4&       1.685&\\
      2798.7&       0.078&\\
      2798.8&       0.385&\\ 
      2803.5&       1.000&\\
            2797.9&   &    1.000\\
            &\\
      2929.5&       0.585&\\
      2937.4&       1.163&\\
            2934.7&   &    0.494\\
            &\\
      9220.8&       0.280&\\
      9246.8&       0.276&\\
            9233.6&     &  0.157\\
            &\\
     10917.2\phantom{ }&       0.099&\\
     10918.3\phantom{ }&       0.011&\\
     10954.8\phantom{ }&       0.108&\\ 
          10935.9\phantom{ }&   &    0.061\\
     
 \hline
\end{tabular}
\label{tab:mgii_fl}
\end{table}

A requirement for the fluorescence mechanism to be important would be that 1) the $\lambda 1240$ transition is optically thick and that 2) there is enough flux at this wavelength to power the observed lines. In the lower panel of Fig.~\ref{fig:endep} we show the optical depths of the Mg II $\lambda 1239.9$ line\footnote{The ratio between the $\lambda \lambda 1239.9 / 1240.4$ lines is fixed by the transition rates to $\tau(\lambda 1239.9 )/\tau(\lambda 1240.4 )=2.66$.} as the dashed lines. The Mg~{\sc ii} $\lambda \lambda 1239.9, 1240.4$ lines are optically thick inside $\sim 2600 $~\kms. This corresponds to the region with enriched Mg abundance and high density from the O/Ne/Mg core. Note that this ejecta model is a result of a spherical average with clumps from the O/Ne/Mg core mixed with H and He rich gas \citep{utrobin21}. Therefore, clumps from the core will have considerably higher optical depths within this region and may also be optically thick at higher velocities. Photons with a comoving wavelength corresponding to these transitions will therefore feed the fluorescence cascade. As the optical depth is sensitive to the Mg abundance, one may expect this to be important mainly for processed material in the core, not in e.g. the hydrogen dominated interaction in Type IIn SNe unless the density is very high.

As for the second condition, the velocity difference between Ly$\alpha$ and the Mg~{\sc ii}~$\lambda \lambda 1239.9, 1240.4$ doublet is $\sim 5900$~\kms. With a velocity width of $\sim 8500$~\kms \ the RS flux from both Ly$\alpha$ and N~{\sc v} $\lambda\lambda 1239, 1243$ can be resonance scattered by the Mg~{\sc ii} $\lambda \lambda 1239.9, 1240.4$ lines. This required velocity difference, combined with a non-dominant fluorescence contribution, should result in a $\lambda\lambda2796,2804$ distribution similar to, but slightly more edge-brightened than H$\alpha$ (as a proxy of Ly$\alpha$), which indeed seems to be the case. From the COS spectrum, we estimate that the Ly$\alpha$ flux from the RS is $\sim 9 \times 10^{-12}$~erg~s$^{-1}$~cm$^{-2}$, and from N~{\sc v} $\lambda \lambda 1239, 1243, \sim 7 \times 10^{-13}$~erg~s$^{-1}$~cm$^{-2}$. In 2018 the total Mg~{\sc ii} fluxes, corrected for Galactic and LMC reddening, were $F(\lambda 2800) \approx 1.1\times 10^{-13}$~erg~s$^{-1}$~cm$^{-2}$ (most of which may be excited thermally) and $F(\lambda 9230) \approx 1.3\times 10^{-14} $~erg~s$^{-1}$~cm$^{-2}$. Therefore, $\lesssim 1 \%$ of the flux from the RS needs to be absorbed by the Mg~{\sc ii} ions in order to feed the cascade, and the effect on the line profiles of Ly$\alpha$ and N~{\sc v} will thus be minimal.
In addition, Ly$\alpha$ from the inner ejecta may contribute. The flux of this component is unknown due to strong interstellar absorption. However, because the velocity width (from H$\alpha$) is only $\sim 3500$~\kms~(Fig.~\ref{fig:total_ejecta_lines}), emission from only a small fraction of the core will have high enough relative velocity to be scattered by the Mg~{\sc ii} ions. It is therefore likely that the RS dominates the {\it fluorescent} Mg~{\sc ii} excitation.

One might argue against the fluorescence scenario that, while the Ly$\alpha$ and  N~{\sc v} $\lambda\lambda 1239, 1243$ from the RS have decreased by a factor $1.5 - 2$ from 2011 to 2017 (Fransson et al., in prep.), the Mg~{\sc ii}~$\lambda\lambda9221,9247$ flux has simultaneously increased by a factor of $\sim 2 $ (Fig.~\ref{fig:linefluxes}). There are, however, several factors why a direct correlation between the total flux at $\sim 1240$ \AA \ and the fluorescence flux is not expected. 

The relative velocity of the fluorescence source and the Mg~{\sc ii} emitting region corresponds to the aforementioned $\sim$5900~\kms \ for the line centre of  Ly$\alpha$. Because the optically thick  Mg~{\sc ii} is located at $\la 2000$~\kms \ (Fig.~\ref{fig:endep}) the RS has to be at $\la 7900$~\kms~for a resonance to take place. The velocity of the RS in the plane of the ER is $\sim 0.8 \ R_{\rm ER}/t \approx 15,500 \ (10 \ {\rm yrs}/t)$~\kms. Therefore, from a rough calculation, fluorescence by RS photons takes place only after $\sim$20 years ($\sim$2007), and strengthens until $\sim$28 years ($\sim$2015), when the surface of a sphere centred on an RS element intersects the largest area inside the core. This calculation assumes a sharp  Ly$\alpha$ line, while the radiative damping will make it broader, hastening the onset of fluorescence. There may also be a contribution from N~{\sc v} $\lambda\lambda 1239, 1243$ emission from the ER (for ejecta with relative velocity close to zero), as well as two-photon emission, but the Ly$\alpha$ contribution is expected to dominate after $\sim$20 years (Fig. \ref{fig:total_ejecta}).

A further effect is that the solid angle of the optically thick Mg~{\sc ii}~$\lambda \lambda 1239.9, 1240.4$ region is increasing with time as the ejecta expands (Sect.~\ref{sec:xray_dep}, Fig.~\ref{fig:endep}) and an increasing fraction of the RS flux may therefore be scattered in these lines. A more detailed calculation 
is out of the scope of this paper.

\subsection{Excitation of the molecular hydrogen}
\label{sec:h2}

\citet{fransson16} reported the discovery of infrared H$_2$ molecular lines. The power source suggested for these lines was $^{44}$Ti decay, as the lines were more centrally concentrated than the edge-brightened H$\alpha$ emission (and indeed X-rays from the ejecta-ER interaction could dissociate the molecules) -- this is also supported by a lack of time evolution in the infrared H$_2$ flux \citep[][]{larsson19b}. The possible excitation mechanisms not related to X-rays from interaction are non-thermal electrons, from cascades caused by positrons from the $^{44}$Ti decay; and UV continuum fluorescence by photons generated internally within the ejecta. Based on comparisons of these excitation models to the NIR spectra, \citet{fransson16} marginally favored UV fluorescence.

Both mechanisms also result in a series of FUV H$_2$ emission lines between 1200 and 1650~\AA. UV continuum fluorescence, in particular, should create a strong peak at 1608~\AA \ \citep{sternberg89,cmc95}. Excitation by non-thermal electrons could create a broader double peak at 1578 and $1608$~\AA \ \citep{dols00,gustin09,france10b}, but a double-peaked feature is also expected in some UV continuum fluorescence models \citep{witt89,france05}. We, however, do not detect any such features in the COS spectrum (see Sect.~\ref{sec:spec}), and are thus unable to distinguish between these mechanisms. The absorption line at 1608~\AA, while not concealing an ejecta H$_2$ emission line with FWHM $\sim2300$~\kms \ per se \citep{fransson16}, none the less makes determining an upper limit for it difficult. This does not apply to a second peak at $1578$~\AA, where we can set a 3$\sigma$ upper limit of $<7.1\times10^{-16}$~erg~s$^{-1}$~cm$^{-2}$ between 1571 and 1584~\AA \ (corresponding to a width of 2500~\kms). To our knowledge, there have not been sufficient modeling studies of the UV line fluxes in relation to the NIR lines, and this limit therefore cannot yet be used to constrain the excitation mechanism.  

\subsection{The outer spots and the high-velocity clump}
\label{sec:spotveloc}

Our spectroscopic observations allow us to examine, for the first time, a trio of recently created outer hotspots south of the ER, whose emergence was described by \citet{fransson15} and \citet[][]{larsson19}; these spots were suggested to be due to slow shocks created by interaction between the fastest ejecta and the clumpy, slowly out-flowing matter between the ER and the southern OR. Their emission (Fig.~\ref{fig:spotspec}) is dominated by Balmer lines, unresolved by STIS, at almost zero line-of-sight velocity ($\sim$100~km~s$^{-1}$), supporting this association. The narrow line profile supports the scenario of slow shocks driven into dense clumps of matter. The lower peak velocity compared to the ER emission, which peaks at a few hundred~km~s$^{-1}$, is in turn consistent with high-latitude material, i.e. low line-of-sight velocities -- as opposed to an extension of the ER along the equatorial plane, which would peak at velocities at least equal to the ER component. In addition to the Balmer lines, we see an [O~{\sc iii}]~$\lambda5008$ feature. While the feature peaks at a velocity consistent with the ER emission, it likely includes a contribution from spot B (see Fig.~\ref{fig:ha_img_zoom}), which was detected in narrow-band imaging at this line by \citet{larsson19}.

An origin in high-latitude material can imply a clumpy BSG wind -- clumping is theoretically expected in the winds of hot stars, and can considerably lower mass loss from the standard prescriptions used in stellar evolution \citep[see e.g.][]{smith14} -- and/or (more likely) instabilities at the interface between the winds launched in the RSG and BSG phases \citep{lb91,ma95}. Regardless, their existence and suggested location highlights the presence of matter connecting the ER and the ORs as discussed by \citet[][]{larsson19}, providing a constraint for models of mass loss in the progenitor system of SN 1987A and other similar SNe. The binary merger model of \citet{mp09} shows no evidence of such matter, for example, while it is not included in the simulations by \citet{orlando15,orlando20}.

A clump of emission seen in the spectra at Position 1, at roughly 1500~km~s$^{-1}$, seems at first glance to correspond to spot B. This association is, however, likely coincidental, as described in Sect.~\ref{sec:spots}: in a combination of fortune and misfortune, our erroneously placed slit at Position 2 overlaps with Position 1 enough that both slits cover spot B; however, the feature at 1500~km~s$^{-1}$ is only seen in Position 1. Furthermore, this velocity is highly unlikely to be associated with any of these spots, as it is a factor of $\sim50$ faster than the material that likely forms the spots between the ER and the ORs \citep{crotts00,larsson19} -- and projection effects should significantly increase this discrepancy. This also suggests the association of the red shoulder with the declination of spot B is coincidental. 

Since the feature is not seen in Position 2, which overlaps with the western half of Position 1, we can narrow down the location of its origin: it is likely that the source of the 1500~km~s$^{-1}$ H$\alpha$ shoulder is located slightly east of spot B. An obvious explanation to consider is emission from the RS: the bipolar schematic picture presented by \citet{france15} in their Fig.~5 suggests that some material crossing the RS in the ejecta south of the ER has a line-of-sight velocity of 1000--2000~km~s$^{-1}$, a clump of which can be responsible for the 1500 km s$^{-1}$ feature (see Fig. \ref{fig:sketch}). In addition, the area where emission from such material is seen (region 'B-r' in their nomenclature) mostly covers the east side of the ER -- and its western edge is close to our Position 2 slit. This feature is not seen in the 2D spectrum at Position 2, but the ER emission is significantly brighter there than in Position 1, and we do see a tentative shoulder in the north-south slice of Position 2 in the lower panel of Fig.~\ref{fig:spotdist}, corresponding to $\sim$1500~km~s$^{-1}$. There is no obvious explanation for the equivalent, fainter feature just north of the ER, but the time difference between the observations reported here and those in \citet{france15} is roughly 3.5 years, in which time the shape of the RS may have slightly changed. We encourage further observations of this region, preferably with integral field spectroscopy.

\section{Summary and future evolution}
\label{sec:concl}

We have examined spectra of the inner ejecta of SN~1987A from 2018, observed using the \emph{HST} in the 1150--10000~\AA~range, and compared them to earlier spectra. We have also created velocity-space maps of the ejecta, which we have used to compare the structures of strong emission lines against each other in 2018 and against earlier velocity maps. This allows us to study the morphology and evolution of emission lines from the ejecta. We have also analysed the spectrum on the south side of the ER, searching for emission from recently-emerged outer spots. The conclusions of this analysis are listed below. See also Fig. \ref{fig:sketch} for a summary of the different emission regions and power sources of the SN 1987A system. 
\begin{itemize}
    \item The H$\alpha$, Mg~{\sc ii} and O~{\sc i} lines show broadly similar edge-brightened spatial distributions, albeit slightly blueshifted in the case of Mg~{\sc ii}. The H$\alpha$, O~{\sc i} and Ca~{\sc ii} lines are consistent with each other in terms of their increasing flux, while the Mg~{\sc ii}~$\lambda\lambda2796, 2804$ and $\lambda\lambda9221, 9247$ doublets have brightened relatively much more after 1999 (by a factor of $\sim$9, as opposed to e.g. H$\alpha$ fluxes that were similar in 1999 and 2018).
    \item Calculations of the X-ray deposition with realistic models of the SN~1987A ejecta and the X-ray spectrum show that X-rays from the CSM interaction shocks deposit their energy throughout the hydrogen envelope, and can penetrate down to the metal rich regions of the core of the SN. X-rays as the primary power source can thus explain the spatial and temporal characteristics we observe from the emission lines. In particular, the large increase in the Mg~{\sc ii} $\lambda\lambda2796,2804$ lines, which account for a large fraction of the cooling, is consistent with this -- and with the importance of Mg~{\sc ii} in the cooling of interacting SNe. 
    \item Measured line flux ratios of different Mg~{\sc ii} lines relative to Mg~{\sc ii}~$\lambda\lambda2796, 2804$ are roughly one third of the predicted ratios from Ly$\alpha$ fluorescence. The strong Mg~{\sc ii}~$\lambda\lambda2796, 2804$ resonance lines are therefore likely powered by a combination of thermal excitation and (to a lesser extent) fluorescence, while the NIR Mg~{\sc ii} lines are excited by fluorescence only; the latter's presence provides a probe of the far-UV radiation below 1240~\AA.
    \item H$_2$ emission lines from the ejecta are not detected in our COS spectrum. We therefore cannot use these lines to distinguish between UV fluorescence or non-thermal electron excitation of NIR H$_2$ lines.  
    \item We detect line emission from three new hotspots outside the ER, previously only observed through imaging \citep{fransson15,larsson19}. The spectrum is dominated by narrow (unresolved, $\leq$670~km~s$^{-1}$) Balmer lines at a line-of-sight velocity consistent with zero, supporting the association of these spots with slow shocks in dense clumps in slowly outflowing CSM outside the ER. We also see an emission feature at $\sim$1500~km~s$^{-1}$, but conclude that this most likely originates at a part of the RS projected south of the ER. Spectroscopic observations of the area to the east of the spots can shed more light on this.
\end{itemize}

We expect the evolving SNR to change in multiple ways in the next several years. The X-ray deposition will continue to shift toward lower velocities as the optical depth decreases, resulting in more emission from metal ions and more information on the structure of the core. More emission from hotspots and high-latitude matter is appearing outside the ER, yielding information on the mass loss history including possible clumping in the winds. At the same time the inner strongly emitting ejecta will reach the RS and, soon afterwards, the ER; the velocity corresponding to the RS in the plane of the ER is decreasing and will reach $\sim4000$~km~s$^{-1}$ in 2023. The interaction between the inner ejecta and the ER will allow us to better probe the density profile of the inner ejecta. Thus we are about to enter a new phase for this unique event, which continues to provide insights into the evolution of SNRs.

\section*{Acknowledgements}

We thank the anonymous referee for their suggestions for improving the paper. We also thank Peter Garnavich, Bruno Leibundgut, Kevin Heng and Jason Spyromilio for helpful suggestions and comments. 

This work is supported by the Swedish National Space Agency and the Swedish Research Council.

This work is based on observations made with the NASA/ESA \textit{Hubble Space Telescope} (programme GO 14753, PI Fransson), obtained through the data archive at the Space Telescope Science Institute (STScI). STScI is operated by the Association of Universities for Research in Astronomy, Inc. under NASA contract NAS 5-26555. The ground-based observations were collected at the European Organization for Astronomical Research in the Southern Hemisphere, Chile(ESO Program 100.D-0-705(A)).

We have made use of the CHIANTI atomic database. CHIANTI is a collaborative project involving George Mason University, the University of Michigan (USA), University of Cambridge (UK) and NASA Goddard Space Flight Center (USA).

\section*{Data availability}

The raw and processed \emph{HST} data used here are publicly available for download at the MAST Portal (\url{https://mast.stsci.edu/portal/Mashup/Clients/Mast/Portal.html}). The VLT/UVES data are publicly available for download at the ESO Archive Science Portal (\url{http://archive.eso.org/scienceportal/home}).

The derived data generated in this research will be shared on reasonable request to the corresponding author.





\bibliographystyle{mnras}
\bibliography{main} 




\bsp	
\label{lastpage}
\end{document}